\documentclass[12pt]{article}

\parskip=\medskipamount

\usepackage{tensor}
\usepackage{ulem}
\usepackage{physics}
\usepackage{amsmath,amssymb,calc, amsthm,bbm, epsfig,psfrag, mathtools}
\usepackage{latexsym,bm,amsfonts}
\usepackage{graphicx, enumerate}
\usepackage{enumitem}
 \usepackage[all,cmtip]{xy}
\usepackage[numbers,sort&compress]{natbib}
\usepackage[dvipsnames]{xcolor}
\usepackage{mathrsfs}
\usepackage{tikz,tikz-cd}

\numberwithin{equation}{section}

\allowdisplaybreaks

\newcommand{\ul}{\underline}

\newcommand{\Comment}[1]{{}}
\definecolor{darkblue}{rgb}{0.15,0.35,0.55}
\definecolor{reddish}{rgb}{0.65, 0.2, 0.2}
\usepackage[linktocpage=true]{hyperref}
\hypersetup{
colorlinks=true,
citecolor=darkblue,
linkcolor=reddish,
urlcolor=darkblue,
pdfauthor={},
pdftitle={},
pdfsubject={}
}

\makeatletter
\renewcommand\section{\@startsection {section}{1}{\z@}%
                                   {-3.5ex \@plus -1ex \@minus -.2ex}
                                   {2.3ex \@plus.2ex}%
                                   {\normalfont\large\bfseries}}
\renewcommand\subsection{\@startsection{subsection}{2}{\z@}%
                                     {-3.25ex\@plus -1ex \@minus -.2ex}%
                                     {1.5ex \@plus .2ex}%
                                     {\normalfont\bfseries}}
\makeatother

\newfont{\goth}{ygoth.tfm scaled 1200}                   

\newcommand{\overbar}[1]{\mkern 1.5mu\overline{\mkern-1.5mu#1\mkern-1.5mu}\mkern 1.5mu}
\newcommand{\Dbar}{\overbar{D}}
\newcommand{\psib}{\overbar{\psi}}

\newcommand{\Qbar}{\overbar{Q}}
\newcommand{\thetab}{\overbar{\theta}}
\def\TT{{T\overbar{T}}}

\def\cbar{{\overbar{c}}}

\def\cEb{{\overbar{\mathcal{E}}}}
\def\cE{{\mathcal{E}}}

\begin{document}
\begin{titlepage}
\begin{flushright}
\today
\end{flushright}
\vspace{5mm}

\begin{center}
{\Large \bf 
ModMax Oscillators and Root-$\TT$-Like Flows in Supersymmetric Quantum Mechanics
}
\end{center}

\begin{center}

{\bf
Christian Ferko${}^{a}$
and
Alisha Gupta${}^{b}$
} \\
\vspace{5mm}

\footnotesize{
${}^{a}$
{\it 
Center for Quantum Mathematics and Physics (QMAP), 
\\ Department of Physics \& Astronomy,  University of California, Davis, CA 95616, USA
}
 \\~\\
${}^{b}$
{\it 
The Academy for Mathematics, Science, and Engineering \\ Morris Hills High School, Rockaway, NJ 07866, USA
}}
\vspace{2mm}
~\\
\texttt{caferko@ucdavis.edu, alishag0101@gmail.com}\\
\vspace{2mm}

\end{center}

\begin{abstract}
\baselineskip=14pt

\noindent We construct a deformation of any $(0+1)$-dimensional theory of $N$ bosons with $SO(N)$ symmetry which is driven by a function of conserved quantities that resembles the root-$\TT$ operator of $2d$ quantum field theories. In the special case of $N=2$ bosons and a harmonic oscillator potential, the solution to the flow equation is the ModMax oscillator of \href{https://arxiv.org/abs/2209.06296}{2209.06296}.
We argue that the deforming operator is related, in a particular special regime, to the dimensional reduction of the $2d$ root-$\TT$ operator on a spatial circle.
It follows that the ModMax oscillator is a dimensional reduction of the $4d$ ModMax theory to quantum mechanics, justifying the name.
We then show how to construct a manifestly supersymmetric extension of this root-$\TT$-like operator for any $(0+1)$-dimensional theory with $SO(N)$ symmetry and $\mathcal{N}=2$ supersymmetry by defining a flow equation directly in superspace. 

\end{abstract}
\vspace{5mm}

\vfill
\end{titlepage}

\newpage
\renewcommand{\thefootnote}{\arabic{footnote}}
\setcounter{footnote}{0}

\tableofcontents{}
\vspace{1cm}
\bigskip\hrule


\allowdisplaybreaks

\section{Introduction}\label{sec:intro}

Whenever one would like to understand some new aspect of quantum field theory, it is useful to find the simplest toy model in which this aspect appears. One way of achieving such a simplification is by reducing the spacetime dimension. As a famous example, several interesting features of supersymmetric gauge theories can already be studied in the setting of supersymmetric quantum mechanics (SUSY-QM). As a $(0+1)$-dimensional theory, SUSY-QM is dramatically simpler than field theories with spatial dimensions. But despite this simplification, one can already study phenomena such as spontaneous supersymmetry breaking -- which can be probed by computing index-like quantities -- in the quantum mechanical setting \cite{WITTEN1982253}.

In this work, we will apply this paradigm to analyze a new class of field theories which contain non-analytic square root expressions in their Lagrangians. The first example of such a theory to be discovered is the Modified Maxwell or ``ModMax'' theory in four spacetime dimensions \cite{Bandos:2020jsw}. The ModMax theory is described by the Lagrangian
\begin{align}\label{modmax_lagrangian}
    \mathcal{L}_{\text{ModMax}} = 
    - \frac{1}{4} \cosh ( \gamma ) F^{\mu \nu} F_{\mu \nu}
    + \frac{1}{4} \sinh ( \gamma ) 
    \sqrt{ 
    \left( F^{\mu \nu} F_{\mu \nu} \right)^2 + \left( F^{\mu \nu} \widetilde{F}_{\mu \nu} \right)^2
    } \, ,
\end{align}
where $\widetilde{F}^{\mu \nu} = \frac{1}{2} \epsilon^{\mu \nu \rho \sigma}F_{\rho\sigma}$ is the Hodge dual of the field strength $F_{\mu \nu}$. The Lagrangian (\ref{modmax_lagrangian}) defines a classical theory of non-linear electrodynamics which is the unique conformally invariant and electric-magnetic duality invariant extension of the Maxwell theory in $d = 4$. It has many interesting features, including plane wave solutions which are well-behaved when $\gamma > 0$ (although the theory allows superluminal propagation when $\gamma < 0$) and which exhibit birefringence. The ModMax theory can be supersymmetrized \cite{Bandos:2021rqy,Kuzenko:2021cvx} and lifts to a $6d$ ModMax-like tensor theory \cite{Bandos:2020hgy}. Progress towards a brane realization of ModMax was presented in \cite{Nastase:2021uvc}. For a pedagogical discussion of theories of non-linear electrodynamics, including the ModMax theory, we refer the reader to \cite{Sorokin:2021tge} and references therein.

Another intriguing observation concerning ModMax is that it arises from a marginal stress tensor deformation of the Maxwell theory \cite{Babaei-Aghbolagh:2022uij,Ferko:2022iru,Conti:2022egv,Ferko:2023ruw}. The Lagrangian (\ref{modmax_lagrangian}) obeys
%
\begin{align}\label{modmax_flow_eqn}
    \frac{\partial \mathcal{L}_{\text{ModMax}}}{\partial \gamma} = \frac{1}{2} \sqrt{ T^{(\gamma) \mu \nu} T_{\mu \nu}^{(\gamma)} - \frac{1}{4} \left( \tensor{T}{^{(\gamma)}^\mu_\mu} \right)^2 } \, , 
\end{align}
where
\begin{align}
    T_{\mu \nu}^{(\gamma)} = g_{\mu \nu} \mathcal{L}_{\text{ModMax}} - 2 \frac{\partial \mathcal{L}_{\text{ModMax}}}{\partial g^{\mu \nu}} \, , 
\end{align}
is the Hilbert stress tensor associated with the ModMax theory at parameter $\gamma$. 
In fact, this flow can be extended to a two-parameter family of commuting deformations whose solution is the ModMax-Born-Infeld theory parameterized by two quantities $\gamma$ and $\lambda$:
\begin{align}\label{modmax_BI_lagrangian}
    \mathcal{L}_{\gamma\text{BI}} = \frac{1}{\lambda} \left( 1 - \sqrt{ 1 - 2 \lambda \left( \cosh ( \gamma ) S + \sinh ( \gamma ) \sqrt{S^2 + P^2} \right) - \lambda^2 P^2 } \right) \, .
\end{align}
Here we have defined $S = - \frac{1}{4} F_{\mu \nu} F^{\mu \nu}$ and $P = - \frac{1}{4} F_{\mu \nu} \widetilde{F}^{\mu \nu}$, which are the two independent Lorentz scalars that can be constructed from the field strength $F_{\mu \nu}$ of an Abelian gauge theory in four spacetime dimensions. The ModMax-Born-Infeld Lagrangian (\ref{modmax_BI_lagrangian}), which was first obtained in \cite{Bandos:2020hgy}, satisfies the two commuting flow equations \cite{Conti:2018jho,Babaei-Aghbolagh:2022uij,Ferko:2022iru,Conti:2022egv,Ferko:2023ruw}
\begin{align}\label{ModMax_BI_two_commuting_flows}
    \frac{\partial \mathcal{L}_{\gamma \text{BI}}}{\partial \lambda} = O_{T^2} \, , \qquad \frac{\partial \mathcal{L}_{\gamma \text{BI}} }{\partial \gamma} = R \, ,
\end{align}
where
\begin{align}
    O_{T^2} = \frac{1}{8} \left( T^{\mu \nu} T_{\mu \nu} - \frac{1}{2} \left( \tensor{T}{^\mu_\mu} \right)^2 \right) \, , \qquad R = \frac{1}{2} \sqrt{ T^{\mu \nu} T_{\mu \nu} - \frac{1}{4} \left( \tensor{T}{^\mu_\mu} \right)^2 } \, .
\end{align}
Furthermore, both of these flows can be made manifestly supersymmetric \cite{Ferko:2022iru,Ferko:2023ruw}, yielding the supersymmetric extension of ModMax-BI which was presented in \cite{Lechner:2022qhb}.

This surprising relationship between theories of non-linear electrodynamics and stress tensor flows is especially interesting because much research activity has been devoted to deformations of quantum field theories by functions of the energy-momentum tensor in the past several years. The most famous example within this class is the $\TT$ deformation of $2d$ QFTs \cite{Zamolodchikov:2004ce,Smirnov:2016lqw,Cavaglia:2016oda}, which is the two-dimensional analogue of the first of the two commuting flow equations in equation (\ref{ModMax_BI_two_commuting_flows}). The $\TT$ deformation defines a one-parameter family of theories $S_\lambda$ which satisfy the differential equation
\begin{align}\label{2d_TT_defn}
    \frac{\partial S_\lambda}{\partial \lambda} = \frac{1}{2} \int d^2 x \, \left( T^{(\lambda) \mu \nu} T^{(\lambda)}_{\mu \nu} - \left( \tensor{T}{^{(\lambda)}^\mu_\mu} \right)^2 \right) \, .
\end{align}
As in equation (\ref{modmax_flow_eqn}), the stress tensor $T_{\mu \nu}^{(\lambda)}$ in (\ref{2d_TT_defn}) carries a $\lambda$ superscript to emphasize that it must be recomputed from the deformed theory $S_\lambda$ at each step along the flow. 

This $\TT$ flow exhibits several remarkable properties. The first is that, although we have introduced this deformation at the level of a classical flow for the Lagrangian, the combination of stress tensors appearing on the right side of (\ref{2d_TT_defn}) actually defines a local operator in the spectrum of any $2d$ quantum field theory. Therefore the $\TT$ flow can be unambiguously defined at the quantum level. Because this quadratic combination of stress tensors has mass dimension $4$, and is therefore irrelevant (in the Wilsonian sense) for a $2d$ field theory, one might expect that deforming by such an operator leads to a loss of analytic control. However, it turns out that despite being irrelevant, this deformation is actually \emph{solvable} in the sense that one can compute many physical observables in the deformed theory in terms of data of the seed theory $S_0$. One example is that the spectrum of the $\TT$-deformed theory on a cylinder of radius $R$ obeys the inviscid Burgers' equation,
\begin{align}\label{inviscid_burgers}
    \frac{\partial E_n}{\partial \lambda} = E_n \frac{\partial E_n}{\partial R} + \frac{P_n^2}{R} \, , 
\end{align}
where $E_n$ and $P_n$ are the energy and momentum, respectively, of a given eigenstate. If the undeformed theory is a CFT, the differential equation (\ref{inviscid_burgers}) can be solved to find
\begin{align}
    E_n ( R , \lambda ) = \frac{R}{2 \lambda} \left( \sqrt{ 1 + \frac{4 \lambda E_n^{(0)} }{R} + \frac{4 \lambda^2 P_n^2}{R^2} } - 1  \right) \, .
\end{align}
Another example is the $\TT$ flow equation for the classical Lagrangian, which can also often be solved in closed form. For instance, beginning from a seed theory $\mathcal{L}_0 = \frac{1}{2} \partial^\mu \phi \partial_\mu \phi$ and solving the flow equation (\ref{2d_TT_defn}), one arrives at
\begin{align}\label{nambu_goto}
    \mathcal{L}_\lambda = \frac{1}{2 \lambda} \left( 1 - \sqrt{ 1 - 2 \lambda \partial^\mu \phi \partial_\mu \phi } \right) \, , 
\end{align}
which is the Lagrangian for the Nambu-Goto string in static gauge with three dimensional target space \cite{Cavaglia:2016oda}. Several other works have developed methods for solving classical $\TT$ flow equations to find the deformed theories given different initial conditions and to study properties of the resulting theories \cite{Bonelli:2018kik,Conti:2018tca,Frolov:2019nrr,Coleman:2019dvf,Frolov:2019xzi,Tolley:2019nmm, Sfondrini:2019smd,Jorjadze:2020ili,Brennan:2019azg,Ferko:2022dpg}. In particular, the $\TT$ deformation appears to preserve desirable features of the seed theory, such as integrability \cite{Smirnov:2016lqw,Chen:2021aid}. Although it would be impractical to summarize the many other interesting results on $\TT$ here, we refer the reader to \cite{Jiang:2019epa} for a review of other aspects of the $\TT$ deformation.

In the present work, we will be primarily interested in another stress tensor deformation, which is the analogue of the second flow equation in (\ref{ModMax_BI_two_commuting_flows}). This other deformation is driven by the recently introduced root-$\TT$ operator \cite{Ferko:2022cix}. In the context of two-dimensional field theories, the root-$\TT$ flow for the action is
\begin{align}\label{root_TT_action_flow}
    \frac{\partial S_\gamma}{\partial \gamma} = \int d^2 x \, \sqrt{ \frac{1}{2}  T^{(\gamma) \mu \nu} T_{\mu \nu}^{(\gamma)} - \frac{1}{4} \left( \tensor{T}{^{(\gamma)}^\mu_\mu} \right)^2 } \, , 
\end{align}
which is of the same form as the operator appearing in the flow equation (\ref{modmax_flow_eqn}) for the ModMax theory. This root-$\TT$ flow appears to share some, but not all, of the interesting features of the $\TT$ deformation.\footnote{See also \cite{Rodriguez:2021tcz,Bagchi:2022nvj,Tempo:2022ndz,Hou:2022csf,Ferko:2023sps} for other work related to root-$\TT$.} For instance, the root-$\TT$ flow preserves classical integrability in many examples \cite{Borsato:2022tmu}. However, it is not known whether one can define a local root-$\TT$ operator at the quantum level, as one can do for the $\TT$ operator. This obstruction prevents us from studying quantum observables, such as the finite-volume spectrum, in a root-$\TT$ deformed theory, although a candidate flow equation for the energy levels,
\begin{align}\label{root_TT_energy_flow}
    \left( \frac{\partial E_n}{\partial \gamma} \right)^2 - \frac{1}{4} \left( E_n - R \frac{\partial E_n}{\partial R} \right)^2 + P_n^2 = 0  \, ,
\end{align}
was proposed in \cite{Ebert:2023tih} and justified using a holographic computation.

Another property which the root-$\TT$ deformation shares with the ordinary $\TT$ flow is that one can often find closed-form solutions for the deformed classical Lagrangian. For instance, beginning with the seed theory of a collection of $N$ free scalar fields $\phi^i$, $i = 1 , \ldots , N$, with the Lagrangian
\begin{align}
    \mathcal{L}_0 = \frac{1}{2} \partial_\mu \phi^i \partial^\mu \phi^i \, , 
\end{align}
the solution to the root-$\TT$ flow equation (\ref{root_TT_action_flow}) is
\begin{align}\label{mod_scalar_theory}
    \mathcal{L}_\gamma = \frac{\cosh ( \gamma )}{2} \partial_\mu \phi^i \partial^\mu \phi^i + \frac{\sinh ( \gamma )}{2} \sqrt{ 2 \partial_\mu \phi^i \partial_\nu \phi^i \partial^\nu \phi^j \partial^\mu \phi^j - \left( \partial_\mu \phi^i \partial^\mu \phi^i \right)^2 } \, .
\end{align}
This theory has been considered in \cite{Ferko:2022cix,Babaei-Aghbolagh:2022leo} and shares exactly the same functional form as the ModMax Lagrangian (\ref{modmax_lagrangian}); in fact, it can be obtained from the ModMax theory by dimensional reduction \cite{Conti:2022egv}. Just as the $4d$ ModMax theory belongs to a two-parameter family of theories which satisfy commuting flow equations driven by a $\TT$-like operator $O_{T^2}$ and a root-$\TT$-like operator $R$, in the $2d$ setting one can consider a family
\begin{align}\label{two_parameter_scalar_family}
    \mathcal{L}_{\gamma, \lambda} = \frac{1}{2 \lambda} \left( 1 - \sqrt{ 1 - 2 \lambda \left( \cosh ( \gamma ) Y_1 + \sinh ( \gamma ) \sqrt{ 2 Y_2 - Y_1^2 } \right) + 2 \lambda^2 \left( Y_1^2 - Y_2 \right) }  \right) \, , 
\end{align}
where for convenience we have defined
\begin{align}\label{intro_scalar_Xi}
    Y_1 = \partial_\mu \phi^i \partial^\mu \phi^i \, , \qquad Y_2 = \partial_\mu \phi^i \partial_\nu \phi^i \partial^\nu \phi^j \partial^\mu \phi^j \, ,
\end{align}
and which also satisfy two commuting flow equations,
\begin{align}\label{ModScalar_two_commuting_flows}
    \frac{\partial \mathcal{L}_{\gamma, \lambda}}{\partial \lambda} = \frac{1}{2} \left( T^{\mu \nu} T_{\mu \nu} - \left( \tensor{T}{^\mu_\mu} \right)^2 \right) \, , \qquad \frac{\partial \mathcal{L}_{\gamma, \lambda} }{\partial \gamma} = \sqrt{ \frac{1}{2}  T^{\mu \nu} T_{\mu \nu} - \frac{1}{4} \left( \tensor{T}{^\mu_\mu} \right)^2 } \, .
\end{align}
The theory (\ref{two_parameter_scalar_family}) is the two-dimensional analogue of the ModMax-BI theory (\ref{modmax_BI_lagrangian}), and it can also be obtained from this $4d$ theory by dimensional reduction. As $\lambda \to 0$, (\ref{two_parameter_scalar_family}) reduces to the ModMax-like theory (\ref{mod_scalar_theory}), which we will sometimes refer to as the Modified Scalar theory. For $\gamma = 0$, (\ref{two_parameter_scalar_family}) becomes the static-gauge Nambu-Goto Lagrangian with an $(N+2)$-dimensional target space, which reproduces (\ref{nambu_goto}) when $N = 1$.

We have now seen two examples, the ModMax theory (\ref{modmax_lagrangian}) and the Modified Scalar theory (\ref{mod_scalar_theory}), which belong to the class of theories defined by non-analytic Lagrangians which we alluded to above. Theories with these square-root-type Lagrangians will be the main focus of this paper. Although they appear to possess some interesting properties, models of this form are difficult to understand for several reasons. One reason is that, unlike the Born-Infeld Lagrangian in $4d$ or the Nambu-Goto Lagrangian in $2d$, the square root appearing in these ModMax-type theories does not admit an expansion in powers of the deformation parameter. In fact, because of the non-analytic structure of the square root in these Lagrangians, we cannot even expand around zero field strength ($F_{\mu \nu} = 0$) or around zero gradients for the scalar fields ($\partial_\mu \phi^i = 0$). This makes it challenging to study these models using perturbation theory.

This brings us back to the initial motivation for this work. In order to better understand these non-analytic ModMax-type or root-$\TT$-like theories, we will seek a quantum mechanical -- i.e. $(0+1)$-dimensional\footnote{One point of possible confusion is that we will sometimes use the term ``quantum mechanics'' (in contrast with ``quantum field theory'') to refer to any $(0+1)$-dimensional model, even when we are discussing purely classical aspects of such theories. In other contexts, we will use words like ``quantum'' (in contrast with ``classical'') to refer to issues related to the quantization of such theories. We trust that the reader can distinguish between these usages based on context.} -- toy model which exhibits the same square-root structure in the Lagrangian. This direction was initiated in \cite{Garcia:2022wad}, where the authors studied a deformed $2d$ oscillator defined by the action
\begin{align}\label{garcia_deformed_oscillator}
    S = \int dt \, \left( \cosh ( \gamma ) L_0 + \sinh ( \gamma ) \sqrt{ E_0^2 - J_0^2 } \right) \, , 
\end{align}
where
\begin{align}
    L_0 = \frac{1}{2} \left( \dot{x}^i \dot{x}^i - x^i x^i \right)  \, , \qquad i = 1 , 2 \, , 
\end{align}
is the Lagrangian for a two-dimensional harmonic oscillator with unit mass and frequency, and $E_0$, $J_0$ are the energy and angular momentum, respectively, of the undeformed theory.

The theory (\ref{garcia_deformed_oscillator}) satisfies a $(0+1)$-dimensional analogue of the root-$\TT$ flow equation which characterizes the ModMax and Modified Scalar theories. In the quantum mechanical example, the conserved quantities appearing in the flow are the energy and total angular momentum, rather than components of the energy-momentum tensor. Furthermore, the model (\ref{garcia_deformed_oscillator}) possesses several other interesting properties such as integrability and a ``deformation map'' which transforms the undeformed harmonic oscillator into the deformed model using a change of variables that depends on kinematical data.\footnote{This map is reminiscent of field-dependent changes of variables which appear in many discussions of the $\TT$ deformation \cite{Conti:2018tca,Guica:2019nzm,Conti:2022egv}.} We will refer to the theory defined by (\ref{garcia_deformed_oscillator}) as the ModMax oscillator.

Because of the rich features of this ModMax oscillator, we view this as an ideal laboratory for exploring properties of non-analytic Lagrangians. In this paper, we will further develop and extend the study of $(0+1)$-dimensional models like (\ref{garcia_deformed_oscillator}) in several directions. Our aim is to find versions of the desirable properties which stress tensor deformations in higher dimensional field theories, such as the $2d$ $\TT$ operator, possess. In particular, the $\TT$ operator is \emph{universal} in that it exists in the spectrum of any translation-invariant field theory; it preserves many \emph{symmetries} of the undeformed theory, such as supersymmetry; and it gives rise to \emph{solvable} flow equations for certain quantities such as the classical Lagrangian. We will argue that some form of each of these three statements is also true for the $(0+1)$-dimensional root-$\TT$-like deformation.

The structure of this paper is as follows. In Section \ref{sec:bosonic}, we define a root-$\TT$-like deformation for \emph{any} quantum mechanical system with $SO(N)$ symmetry, generalizing beyond the known harmonic oscillator case with $N = 2$. In Section \ref{sec:dim_red}, we demonstrate that the $2d$ Modified Scalar theory can be dimensionally reduced to the ModMax oscillator in a particular limit, and that the details of this dimensional reduction procedure explain some of the differences between the $2d$ $\TT$ and root-$\TT$ flows and their $(0+1)$-dimensional analogues. In Section \ref{sec:SUSY}, we extend this class of deformations to theories with fermions by formulating a manifestly supersymmetric version of the root-$\TT$-like flow. Finally, in Section \ref{sec:conclusion} we summarize our results and identify some open questions. We have also relegated one technical computation to Appendix \ref{app:other}.

\section{Bosonic Flows}\label{sec:bosonic}

In this section, we will consider flow equations for the classical Lagrangian of a $(0+1)$-dimensional system without any supersymmetry. The resulting differential equations which we obtain will be related to the manifestly supersymmetric flows considered in Section \ref{sec:SUSY}, although without the added complication of fermions and superfield quantities. More precisely, one can obtain purely bosonic flows of the form considered in this section by beginning with a supersymmetric flow of Section \ref{sec:SUSY}, integrating out the superspace coordinates, setting fermions to zero, and putting any auxiliary fields on-shell.

The flows considered here are inspired by classical stress tensor deformations of field theories in $d \geq 2$ which were reviewed in the introduction. In that setting, one considers a deformation for the Lagrangian $\mathcal{L}$ of the field theory which takes the form
\begin{align}\label{general_T_flow}
    \frac{\partial \mathcal{L}_\lambda}{\partial \lambda} = f \left( T_{\mu \nu}^{(\lambda)} \right) \, , 
\end{align}
where $f \left( T_{\mu \nu}^{(\lambda)} \right)$ is some Lorentz scalar constructed from the energy-momentum tensor $T_{\mu \nu}^{(\lambda)}$ of the Lagrangian $\mathcal{L}_\lambda$. Because every translation-invariant field theory possesses a conserved stress tensor $T_{\mu \nu}$, a flow equation of the form (\ref{general_T_flow}) is ``universal'' in the sense that we can attempt to solve this differential equation with any initial condition $\mathcal{L}_0$ to obtain a one-parameter family of deformed theories.

In the context of a $(0+1)$-dimensional system defined by an action principle
\begin{align}
    S = \int \, dt \, L \, , 
\end{align}
the analogue of the energy-momentum tensor $T_{\mu \nu}$ is the Hamiltonian $H$ of the model. Therefore, one class of universal flow equations that one could consider for quantum-mechanical systems are those driven by functions of the Hamiltonian (or energy),
\begin{align}\label{fH_deformation}
    \frac{\partial H}{\partial \lambda} = f \left( H^{(\lambda)} \right) \, .
\end{align}
Such flows have been studied in \cite{Gross:2019ach,Gross:2019uxi}, motivated in part by dimensional reduction of $f ( T_{\mu \nu} )$ flows from field theories in $d = 2$ spacetime dimensions. Although these Hamiltonian flows are in some sense simple, one can already observe interesting phenomena in the resulting deformed theories. For instance, one can study signatures of chaos in $f(H)$ deformed versions of SYK models \cite{He:2022ryk}, and flows of this type have appeared in the study of deformed string worldsheets in pp wave backgrounds \cite{Nastase:2022dvy}; the latter are connected to $\TT$-like deformations of spin chains, which have also been analyzed in \cite{Pozsgay:2019xak,Marchetto:2019yyt}.

One of our goals in the present work is to generalize flow equations like (\ref{fH_deformation}) to depend on additional conserved charges, which will allow us to analyze a richer class of deformations. It is natural to expect that one must consider flows that depend on at least two independent charges in order to realize some of the behavior that is seen in stress tensor deformations of higher-dimensional field theories. For instance, in a $2d$ QFT, there are two independent Lorentz scalars that can be constructed from the energy-momentum tensor, namely $T^{\mu \nu} T_{\mu \nu}$ and $\tensor{T}{^\mu_\mu}$, and a general flow (\ref{general_T_flow}) can therefore be written as
\begin{align}\label{general_T_flow_parameterized}
    \frac{\partial \mathcal{L}_\lambda}{\partial \lambda} = f \Big( T^{(\lambda) \mu \nu} T_{\mu \nu}^{(\lambda)} \, , \, \tensor{T}{^{(\lambda)}^\mu_\mu} \Big) \, , 
\end{align}
which depends on a function of two variables. To find analogues of these flows in quantum mechanics, we must introduce a second conserved quantity besides the Hamiltonian.\footnote{Deformations of quantum mechanics which mirror the structure of $J \overbar{T}$ deformations \cite{Guica:2017lia,Bzowski:2018pcy,Guica:2019vnb,Anous:2019osb,Guica:2020eab,Guica:2021pzy,Guica:2021fkv} in field theories, and are driven by combinations of both the Hamiltonian $H$ and some other conserved global charge $Q$, have been considered in \cite{Chakraborty:2020xwo}. However, we will be interested in a different class of deformations which gives ModMax-like or root-$\TT$-like structures.}

In the next subsection, we will accomplish this by restricting to theories which possess an additional $SO(N)$ symmetry and an associated set of conserved angular momenta $J_{mn}$. In such models, we can consider flow equations that are driven by a function of both the energy $E$ (or Hamiltonian $H$) and the total angular momentum $J^2 = J_{mn} J^{mn}$,
\begin{align}
    \frac{\partial L_\lambda}{\partial \lambda} = f ( E , J^2 ) \, .
\end{align}
This class of deformations is sufficiently broad to exhibit behavior which is similar to that of other interesting stress tensor flows in two dimensional field theories, such as the root-$\TT$ deformation.

We make one comment about terminology. We will reserve terms like ``$\TT$ deformation'' and ``root-$\TT$ deformation'' for the usual versions of these flows in two-dimensional field theories. The analogous deformations in quantum mechanics will be referred to as ``$\TT$-like'' and ``root-$\TT$-like'' deformations, respectively, or as ``$(0+1)$-dimensional $\TT$'' and ``$(0+1)$-dimensional root-$\TT$'' flows, to distinguish them from the field theory examples. For short, we will sometimes use the phrases ``QM-$\TT$ flow'' and ``QM-root-$\TT$ flow'' to denote these quantum mechanical deformations.

\subsection{General $f ( E , J^2 )$ Deformations}\label{sec:general_f_H_J2}

We consider an action for a collection of bosons $x^i ( t )$, where $i = 1 , \ldots , N$,
\begin{align}
    S = \int \, dt \, L (  x^i , \dot{x}^i  ) \, .
\end{align}
Furthermore, we will assume that the theory is invariant under $SO(N)$ rotations,
\begin{align}
    x^i ( t ) \longrightarrow \tensor{R}{^i_j} x^j ( t ) \, , \qquad R \in SO ( N ) \, .
\end{align}
One example of such a theory is a sigma model for a particle moving on an $N$-dimensional target space with an $SO(N)$-invariant metric $g_{ij}$ and potential $V$:
\begin{align}
    L = g_{ij} ( x ) \dot{x}^i \dot{x}^j - V \left( x \right) \, .
\end{align}
For simplicity, in the remainder of this section we will assume that $g_{ij} = \delta_{ij}$. Because indices are raised and lowered with the trivial flat metric, we need not distinguish between upstairs and downstairs indices.

We will be interested in two conserved currents within this family of models. The first is the total energy
\begin{align}
    E = \frac{\partial L}{\partial \dot{x}^i} \dot{x}^i - L \, , 
\end{align}
which is the conserved Noether current associated with time translations,
\begin{align}
    t \to t + \epsilon \, , \qquad \delta x^i = \epsilon \dot{x}^i \, .
\end{align}
There is also a collection of angular momenta which are the conserved quantities associated with $SO(N)$ rotations of the bosons $x^i$. For each pair of indices $n \neq m$, the generator $\delta^{(nm)}$ of the corresponding symmetry (not to be confused with the Kronecker delta $\delta^{nm}$) acts as
\begin{align}
    \delta^{(nm)} x_i = \epsilon_{ij}^{(nm)} x^j \, , 
\end{align}
where 
\begin{align}\label{epsilon_defn}
    \epsilon_{ij}^{(nm)} = \begin{cases} 
      1 & i = n , j = m \\
      - 1 & i = m , j = n  \\
      0 & \text{otherwise}  
   \end{cases} \, .
\end{align}
The Noether current associated with this transformation is
\begin{align}
    J_{nm} &= \frac{\partial L}{\partial \dot{x}^n} x_m - \frac{\partial L}{\partial \dot{x}^m} x_n \, .
\end{align}
The total angular momentum defined by
\begin{align}
    J^2 = J^{nm} J_{nm} \, ,
\end{align}
is therefore also a conserved quantity.

A general deformation of the Lagrangian by a function of the conserved energy $E$ and total angular momentum $J^2$ can then be written as
\begin{align}\label{general_f_H_Jsq}
    \frac{\partial L_\lambda}{\partial \lambda} = f \left( E , J^2 \right) \, , 
\end{align}
where $f$ is an arbitrary function of two variables.

In order to obtain explicit partial differential equations which we can attempt to solve in closed form, it will be convenient to re-express the quantities appearing in (\ref{general_f_H_Jsq}) in terms of a basis for the ring of $SO(N)$ scalars that can be constructed from $x^i$ and $\dot{x}^i$. Choosing such a basis of scalars is a common first step in studying classical flow equations driven by combinations of conserved currents. For instance, in studying stress tensor deformations of $4d$ gauge theories, one typically chooses a basis including $S = - \frac{1}{4} F_{\mu \nu} F^{\mu \nu}$ and $P = - \frac{1}{4} F_{\mu \nu} \widetilde{F}^{\mu \nu}$, which are the two functionally independent Lorentz scalars that can be constructed from an Abelian field strength in four spacetime dimensions. In the $(0+1)$-dimensional setting, we choose the three $SO(N)$ invariants
\begin{align}\label{QM_Xi_def}
    X_1 = x^i x_i \, , \qquad X_2 = \dot{x}^i \dot{x}_i \, , \qquad X_3 = x^i \dot{x}_i \, .
\end{align}
The three quantities $X_1, X_2, X_3$ are manifestly singlets under $SO(N)$ transformations because all of the $SO(N)$ indices are contracted. Furthermore, we claim that any other $SO(N)$ invariant can be written as a function of these three basis scalars. As a simple example, for the case of $N = 2$ scalars we could also construct the antisymmetric combination
\begin{align}
    X_4 = \epsilon_{ij} \dot{x}^i x^j = \dot{x}^1 x^2 - \dot{x}^2 x^1 \, , 
\end{align}
but one can show that this invariant satisfies
\begin{align}
    X_1 X_2 = X_3^2 + X_4^2 \, .
\end{align}
Therefore the putative new scalar $X_4$ is, in fact, functionally dependent upon our three basis elements $X_1, X_2, X_3$. We now consider a Lagrangian which is a general function of these three independent scalars,
\begin{align}
    L ( x^i, \dot{x}^i ) = L ( X_1, X_2, X_3 ) \, .
\end{align}
It is straightforward to evaluate the two conserved quantities $E$ and $J^2$ in terms of partial derivatives of $L$ with respect to the three $X_i$. The energy is
\begin{align}
    E = 2 X_2 \frac{\partial L}{\partial X_2} + X_3 \frac{\partial L}{\partial X_3} - L \, , 
\end{align}
whereas each component of the angular momentum takes the form
\begin{align}
    J^{nm} &= 2 \frac{\partial L}{\partial X_2} \left( \dot{x}^n x^m - \dot{x}^m x^n \right) \, .
\end{align}
The total angular momentum is then
\begin{align}
    J^2 = J^{nm} J_{nm} = 4 \left( \frac{\partial L}{\partial X_2} \right)^2 \left( X_1 X_2 - X_3^2 \right) \, .
\end{align}
We may therefore write the most general deformation of an $SO(N)$-symmetric Lagrangian by a function of the energy and total angular momentum as
\begin{align}
    \frac{\partial L}{\partial \lambda} = f \left[ 2 X_2 \frac{\partial L}{\partial X_2} + X_3 \frac{\partial L}{\partial X_3} - L \, , \, 4 \left( \frac{\partial L}{\partial X_2} \right)^2 \left( X_1 X_2 - X_3^2 \right) \right] \, ,
\end{align}
where $f$ is an arbitrary function of two variables. In the case where the function $f$ only depends on its first argument (the total energy) but not on the second argument, such deformations are similar to the $f ( H )$ flows considered in \cite{Gross:2019ach,Gross:2019uxi}, although we have written such flows in the Lagrangian formulation rather than as a differential equation for the Hamiltonian. Next we will choose a particular functional form of the function $f$ which is motivated by ModMax-like theories (or root-$\TT$-like deformations) and is amenable to closed-form solution.

\subsection{Root-$\TT$-Like Deformations}

The root-$\TT$ operator in a two-dimensional field theory is defined by the combination
\begin{align}
    R^{(2d)} = \sqrt{ \frac{1}{2} T^{\mu \nu} T_{\mu \nu} - \frac{1}{4} \left( \tensor{T}{^\mu_\mu} \right)^2 } \, .
\end{align}
This object is a square root of a particular combination of bilinears in conserved currents, which in this setting are built from the stress tensor. We would like to study an analogue of this operator in quantum mechanics, which takes the same schematic form but which is built from the bilinears $E^2$ and $J^2$, and thus which falls into the class of $f ( E, J^2 )$ deformations of Section \ref{sec:general_f_H_J2}.

A natural guess for such a root-$\TT$-like operator is
\begin{align}\label{rtt_QM_first_mention}
    R = \sqrt{ E^2 - J^2 } \, .
\end{align}
This combination was first considered in \cite{Garcia:2022wad}, where it was applied to a seed theory of $N = 2$ bosons $x^i$ subject to a harmonic oscillator potential. In this section, we would like to generalize this analysis to deformations of any $SO(N)$ invariant Lagrangian of $N$ bosons. We will see that the generalization from the $N = 2$ harmonic oscillator to arbitrary $N$ is straightforward, and the same functional form of the deformed Lagrangian (involving hyperbolic trigonometric functions) appears. However, perhaps more surprisingly, we will see that the partial differential equation for the Lagrangian driven by (\ref{rtt_QM_first_mention}) also admits a closed-form solution for a broader class of seed theories.

We begin by expressing the operator $R$ in terms of partial derivatives of $L$ with respect to the three scalars $X_i$, following the general results of Section \ref{sec:general_f_H_J2}. The quadratic combination $E^2 - J^2$ can be written as
\begin{align}
    E^2 - J^2 = \left( L - X_3 \frac{\partial L}{\partial X_3} \right)^2 - 4 X_2 \frac{\partial L}{\partial X_2} \left( L - X_3 \frac{\partial L}{\partial X_3} \right) + 4 \left( X_3^2 + X_2^2 - X_1 X_2 \right) \left( \frac{\partial L}{\partial X_2} \right)^2 \, . 
\end{align}
Therefore, the partial differential equation arising from the root-$\TT$-like flow is
\begin{align}\label{squared_root_TT_flow}
    \left( \frac{\partial L}{\partial \gamma} \right)^2 = \left( L - X_3 \frac{\partial L}{\partial X_3} \right)^2 - 4 X_2 \frac{\partial L}{\partial X_2} \left( L - X_3 \frac{\partial L}{\partial X_3} \right) + 4 \left( X_3^2 + X_2^2 - X_1 X_2 \right) \left( \frac{\partial L}{\partial X_2} \right)^2 \, .
\end{align}
It is easy to check by explicit computation that one solution to this flow is
\begin{align}\label{deformed_oscillator_solution}
    L_\gamma = \frac{\cosh ( \gamma )}{2} ( X_2 - X_1 ) + \frac{\sinh ( \gamma )}{2} \sqrt{ 4 X_3^2 + \left( X_2 - X_1 \right)^2 } \, .
\end{align}
This theory arises from deforming the seed Lagrangian
\begin{align}
    L_0 = \frac{1}{2} \dot{x}^i \dot{x}_i - \frac{1}{2} x^i x_i \, , 
\end{align}
which is the theory of $N$ bosons $x^i$ with unit mass, each subject to a harmonic oscillator potential. When $N = 2$, this theory reduces to the deformed oscillator of \cite{Garcia:2022wad}.

We would like to look for other interesting classes of solutions to (\ref{squared_root_TT_flow}). One reason to expect that such solutions might exist is because the $2d$ root-$\TT$ flow is surprisingly simple and universal, at least when deforming theories of $N$ bosons. We briefly recall the logic for the $2d$ case by way of comparison. Given a collection of $N$ scalar fields $\phi^i$, $i = 1 , \ldots , N$, there are two independent derivatively coupled Lorentz scalars
\begin{align}
    Y_1 = \partial_\mu \phi^i \partial^\mu \phi^i \, , \qquad Y_2 = \partial_\mu \phi^i \partial_\nu \phi^i \partial^\nu \phi^j \partial^\mu \phi^j \, ,
\end{align}
as we mentioned around equation (\ref{intro_scalar_Xi}). For a general Lagrangian $\mathcal{L} ( Y_1, Y_2 )$ which depends on these two Lorentz scalars, the $2d$ root-$\TT$ flow equation is
\begin{align}\label{2d_root_TT_Lagrangian_flow}
    \frac{\partial \mathcal{L}_\gamma}{\partial \gamma} = \sqrt{ 2 Y_2 - Y_1^2 } \left( \frac{\partial \mathcal{L}_\gamma}{\partial Y_1} + 2 Y_2 \frac{\partial \mathcal{L}_\gamma}{\partial Y_2} \right) \, .
\end{align}
Although this appears to be a fairly complicated partial differential equation, the general solution is in fact rather simple. We first define scalars $Y_1^{(\gamma)}$ and $Y_2^{(\gamma)}$ by
\begin{align}\label{2d_deformed_scalars}
    Y_1^{(\gamma)} &= \cosh ( \gamma ) Y_1 + \sinh ( \gamma ) \sqrt{ 2 Y_2 - Y_1^2 } \, , \nonumber \\
    Y_2^{(\gamma)} &= \cosh ( 2 \gamma ) Y_2 + \sinh ( 2 \gamma ) Y_1 \sqrt{ 2 Y_2 - Y_1^2 } \, .
\end{align}
Then the solution to the differential equation (\ref{2d_root_TT_Lagrangian_flow}) is
\begin{align}\label{general_2d_solution}
    \mathcal{L}_\gamma ( Y_1, Y_2 ) = \mathcal{L}_0 \left( Y_1^{(\gamma)} \, , \, Y_2^{(\gamma)} \right) \, .
\end{align}
That is, we simply take the initial condition $\mathcal{L}_0$ which defines the undeformed theory, and replace all instances of $Y_1, Y_2$ with $Y_1^{(\gamma)}$, $Y_2^{(\gamma)}$. Since the deformed scalars (\ref{2d_deformed_scalars}) satisfy 
\begin{align}
    Y_1^{(\gamma = 0)} = Y_1 \, , \qquad Y_2^{(\gamma = 0)} = Y_2 \, , 
\end{align}
this solution manifestly reproduces the correct initial condition as $\gamma \to 0$.

We now investigate whether there is a similar structure for the QM-root-$\TT$ flow defined by (\ref{squared_root_TT_flow}). The first obvious difference is that the $(0+1)$-dimensional flow equation features three independent $SO(N)$ scalars $X_1, X_2, X_3$ rather than the two Lorentz scalars $Y_1, Y_2$ in the $d = 2$ case. To simplify, let us assume that the Lagrangian of the $(0+1)$-dimensional theory depends on the first two $SO(N)$ scalars only through the combination
\begin{align}
    X \equiv X_2 - X_1 = \dot{x}^i \dot{x}_i - x^i x_i \, .
\end{align}
This is true, for instance, in the case of the unit-mass harmonic oscillator and the ModMax-like deformed oscillator (\ref{deformed_oscillator_solution}). Making the ansatz
\begin{align}
    L ( X_1, X_2, X_3 ) = L ( X, X_3 )
\end{align}
in the flow equation (\ref{squared_root_TT_flow}), we find that the flow will not close on the two variables $X, X_3$ unless we impose the consistency condition
\begin{align}\label{two_var_consistency}
    X \left( \frac{\partial L}{\partial X} \right)^2 - \frac{\partial L}{\partial X} \left( L - X_3 \frac{\partial L}{\partial X_3} \right) = 0 \, .
\end{align}
The constraint (\ref{two_var_consistency}) will be satisfied if either $\frac{\partial L}{\partial X} = 0$, which we will ignore for the moment, or if the Lagrangian takes the form
\begin{align}
    L = X f \left( \frac{X_3}{X} , \gamma \right) \, , 
\end{align}
where $f$ is an arbitrary function. To ease notation, we define a new variable $Z \equiv \frac{X_3}{X}$. Then the root-$\TT$-like flow (\ref{squared_root_TT_flow}) imposes the following differential equation on $f ( Z, \gamma )$:
\begin{align}
    \left( 1 + 4 Z^2 \right) \left( f  - Z \frac{\partial f}{\partial Z} \right)^2 = \left( \frac{\partial f}{\partial \gamma} \right)^2 \, , 
\end{align}
which is satisfied if
\begin{align}
    f ( Z , \gamma ) = Z g \left( \frac{\cosh ( \gamma )}{Z} + \sinh ( \gamma ) \sqrt{ 4 + \frac{1}{Z^2} } \right) \, , 
\end{align}
and where $g$ is another arbitrary function of one variable.

We have therefore arrived at a general class of solutions to the $(0+1)$-dimensional root-$\TT$ flow equation. Any Lagrangian
\begin{align}\label{general_arctanh_solution}
    L ( \gamma , X_1, X_2, X_3 ) = \frac{X_3}{2} \, g \left( \cosh ( \gamma ) \frac{X_2 - X_1}{X_3} + \sinh ( \gamma ) \sqrt{ 4 + \frac{ \left( X_2 - X_1 \right)^2}{ X_3^2 } } \right)
\end{align}
is a solution to the flow (\ref{squared_root_TT_flow}), where $g$ is an arbitrary function. In particular, choosing this function to be the identity,
\begin{align}
    g ( x ) = x \, , 
\end{align}
we recover the ModMax oscillator of equation (\ref{deformed_oscillator_solution}).

\subsection{Properties of Root-$\TT$-Like Flow}\label{sec:properties_solutions}

We now compare the result of this calculation to the general solution (\ref{general_2d_solution}) for the two-dimensional root-$\TT$ flow. In both settings, we have found infinitely many solutions of the respective flow equation. These solutions are labeled by functions $\mathcal{L}_0 ( Y_1, Y_2 )$ of two variables in the $2d$ case and by functions $g ( x )$ of one variable in the quantum mechanical case. This offers us many new closed-form solutions to the flow equation (\ref{squared_root_TT_flow}) that we have derived above. If this flow equation has similar solvability properties as the two-dimensional $\TT$ deformation -- insofar as we can compute certain observables in the deformed theories in terms of quantities in the undeformed theory at $\gamma = 0$ -- then we can potentially learn about such observables in an infinite family of deformed models.\footnote{This would be in a similar spirit as other solvable deformations of quantum mechanics, such as the one discussed in \cite{Grassi:2018bci}; in that context, one can make exact statements about a model with the Hamiltonian $H = \cosh ( p ) + V_N ( x )$ for any polynomial potential $V_N ( x )$.}

However, the solutions which we have found for the QM-root-$\TT$ flow are different and more restricted than the ones in the two-dimensional context. First note that, for more general functions besides $g(x) = x$, even the undeformed Lagrangian corresponding to the $\gamma \to 0$ limit of (\ref{general_arctanh_solution}) will be ill-defined when $X_3 = 0$. For instance, if $g(x) = x + \epsilon x^2$, then the Lagrangian of the seed theory contains a term proportional to
\begin{align}
    \frac{ ( X_2 - X_1 )^2}{X_3} \, , 
\end{align}
which diverges when $\dot{x}^i x_i = 0$. Physically, this is rather strange; such a term causes the total energy of the particle to grow without bound when the velocity $\dot{x}^i$ is orthogonal to the position $x^i$. In particular, we cannot have a finite Lagrangian for a kinematical configuration in which $x^i \neq 0$ but all velocity components $\dot{x}^i$ are vanishing.

This is very different from the analogous solutions to the $2d$ flow equation. For the $2d$ root-$\TT$ flow, we can easily write down the solution corresponding to an \emph{arbitrary} seed theory using (\ref{general_2d_solution}), and the solutions generically appear to be well-behaved. In contrast, for the $(0+1)$-dimensional root-$\TT$-like deformation, we can only write down a solution if the seed theory takes the form $L_0 = X_3 g \left( \frac{X_2 - X_1}{X_3} \right)$ for some function $g$, and when $g(x) \neq x$ the resulting theories are not well-defined at kinematical points where $X_3 = 0$.

Another difference is that the $(0+1)$-dimensional root-$\TT$-like flow appears to have only a \emph{single} scalar structure which becomes deformed with $\gamma$, as opposed to two Lorentz scalars which are deformed in the $2d$ flow. By this we mean that the general solution (\ref{general_arctanh_solution}) takes the schematic form of a replacement of the undeformed quantity $\frac{X_2 - X_1}{X_3}$ as
\begin{align}
    \frac{X_2 - X_1}{X_3} \longrightarrow \cosh ( \gamma ) \frac{X_2 - X_1}{X_3} + \sinh ( \gamma ) \sqrt{ 4 + \frac{ \left( X_2 - X_1 \right)^2}{ X_3^2 } } \, , 
\end{align}
modulo the caveat that the undeformed Lagrangian must take the form of a product of this quantity with $X_3$. This is somewhat similar to the replacement of the Lorentz scalar $Y_1$ with its deformed version $Y_1^{(\gamma)}$ in the usual root-$\TT$ flow. However, in the $(0+1)$-dimensional flow there is no analogue of the second tensor structure $Y_2$ which is replaced by $Y_2^{(\gamma)}$. In fact, there do not seem to be any other solutions to the QM-root-$\TT$ flow equation (\ref{squared_root_TT_flow}) which are obtained by performing such a simple replacement, except for the following trivial example: if we assume that the Lagrangian is independent of $X_1$ and $X_2$, so that
\begin{align}
    L = L ( \gamma, X_3 ) \, , 
\end{align}
then the flow equation reduces to
\begin{align}
    \left( \frac{\partial L}{\partial \gamma} \right)^2 = \left( L - X_3 \frac{\partial L}{\partial X_3} \right)^2 \, , 
\end{align}
which is solved by any Lagrangian
\begin{align}
    L = e^{\pm \gamma} f \left( e^{\mp \gamma} X_3 \right) \, , 
\end{align}
where $f$ is an arbitrary function. This can be viewed as a re-scaling of the time coordinate as $t \to e^{\pm \gamma} t$. This class of trivial solutions is reminiscent of the solution to the $2d$ root-$\TT$ flow in the case of a single compact boson, which merely re-scales the kinetic term in a similar fashion and can be interpreted as changing the target space radius.

Finally, we point out one additional difference between the root-$\TT$ flow and the QM-root-$\TT$ flow, which concerns commutativity with the appropriate version of $\TT$. It was pointed out in \cite{Ferko:2022cix} that there appears to be a commuting square of deformations by $\TT$ and root-$\TT$ which can be visualized as
\begin{center}
\begin{tikzcd}
    {{S}_0} && {{S}_\gamma} \\
    \\
    {{S}_\lambda} & {} & {{S}_{(\lambda, \gamma)}}
    \arrow["{{R}}", from=1-1, to=1-3]
    \arrow["{{O_\TT}}"', from=1-1, to=3-1]
    \arrow["{{R}}"', from=3-1, to=3-3]
    \arrow["{{O_\TT}}", from=1-3, to=3-3]
\end{tikzcd} \; .
\end{center}
That is -- at least for the classical Lagrangian flows of theories of $N$ bosons -- one can either first deform a seed action $S_0$ by the root-$\TT$ operator $R$, and then use the resulting deformed theory as a seed for a second flow by the $\TT$ operator $O_{\TT}$, or first deform $S_0$ by $\TT$ and then by root-$\TT$, and the resulting actions $S_{(\lambda, \gamma)}$ agree. This might be interpreted as the statement that some notion of ``curvature'' in the space of field theories is vanishing in the two-dimensional subspace traced out by the $\TT$ and root-$\TT$ flows.

This commutativity between the $2d$ $\TT$ and root-$\TT$ deformations appears to persist at the level of the candidate flow equations for the energies, although the root-$\TT$ deformed spectrum has only been justified via circumstantial evidence from holography rather than a fully rigorous quantum definition \cite{Ebert:2023tih}. Commutativity of these flows is a powerful constraint, and in fact one can reverse the logic and \emph{impose} commutativity of a candidate stress tensor deformation with $\TT$ in order to single out root-$\TT$ as the unique marginal flow with this property for $2d$ field theories, under some assumptions.

Therefore it is another important difference that the QM-root-$\TT$ flow defined by equation (\ref{squared_root_TT_flow}) does \emph{not} commute with the appropriate quantum mechanical version of the $\TT$ deformation. It is somewhat cumbersome to see this because we have defined the QM-root-$\TT$ flow in the Lagrangian formulation, whereas the QM-$\TT$ flow is typically defined in terms of the Hamiltonian. For this reason we have relegated the proof that these flows do not commute to Appendix \ref{app:other}. The argument proceeds by exhibiting a single counterexample: namely, we demonstrate that the $(0+1)$-dimensional versions of $\TT$ and root-$\TT$ do not commute when applied to the seed theory which is the $N$-dimensional harmonic oscillator. We show this by computing the two Lagrangians obtained by performing the deformations in either order, perturbatively in the velocities $\dot{x}^i$ (or momenta $p^i$), and showing that the terms in this perturbative expansion disagree. Note that such an expansion is possible for the ModMax oscillator but not for the four-dimensional ModMax theory, which is non-analytic in field strengths near $F_{\mu \nu} = 0$ and thus does not admit a Taylor expansion. In contrast, the Lagrangian (\ref{deformed_oscillator_solution}) actually does possess an analytic expansion near zero velocities, so long as we assume $x^i x_i \neq 0$. This is because the argument of the square root is $E_0^2 - J_0^2$, and the total energy $E_0^2$ contains a contribution from the potential energy which is non-vanishing at zero velocity.

It may not come as a surprise that the properties of the $(0+1)$-dimensional $\TT$ flow are very different from those of its two-dimensional counterpart -- at least when applied to the seed theory of a harmonic oscillator -- 
 because the quantum mechanical $\TT$ flow is obtained from a dimensional reduction of the $2d$ flow for a \emph{conformal} seed theory, and the harmonic oscillator cannot be obtained by compactifying a CFT on a spatial circle and keeping only the zero mode sector. In the next section we will review this Kaluza-Klein reduction of the $\TT$ operator and show that a similar dimensional reduction is possible, in a certain regime, to obtain the QM-root-$\TT$ flow from the $2d$ root-$\TT$ flow.

\section{Dimensional Reduction of $2d$ Flows}\label{sec:dim_red}

We have seen that various properties of the usual $\TT$ and root-$\TT$ flows in two spacetime dimensions -- such as their commutativity -- fail for their $(0+1)$-dimensional analogues. In this section our goal is to understand this via a dimensional reduction argument from $(1+1)$ dimensions to $(0+1)$ dimensions on a spatial circle. The reduction of the $2d$ $\TT$ deformation to the QM-$\TT$ deformation was explained in \cite{Gross:2019ach,Gross:2019uxi} and will be briefly reviewed here. However, the dimensional reduction which yields the quantum mechanical version of root-$\TT$ from the corresponding $2d$ flow is new.

The upshot of our analysis is that, although there exist certain assumptions under which each of the $2d$ $\TT$ and root-$\TT$ flows can be reduced to deformations of quantum mechanics, the sets of assumptions for the two flows are incompatible. That is, there is no regime in the $2d$ field theory under which \emph{both} of the deformations by $\TT$ and root-$\TT$ admit a dimensional reduction to the corresponding flows for a $(0+1)$-dimensional theory. This explains why the QM-$\TT$ and QM-root-$\TT$ flows do not commute, and more generally why they seem to exhibit very different properties than their $2d$ analogues: in order to perform the dimensional reduction, for either deformation, one must impose additional criteria which break some of the properties of the flow.

\subsection{Reduction of $\TT$ Flow}

Consider a one-parameter family of Lagrangians $S_\lambda$ for two-dimensional field theories which satisfy the $\TT$ flow equation (\ref{2d_TT_defn}),
\begin{align}\label{TT_flow_eqn_reduction}
    \frac{\partial S_\lambda}{\partial \lambda} = \frac{1}{2} \int d x \, dt \, \left( T^{(\lambda) \mu \nu} T^{(\lambda)}_{\mu \nu} - \left( \tensor{T}{^{(\lambda)}^\mu_\mu} \right)^2 \right) \, ,
\end{align}
and let the initial condition be $S_0 = \int d^2 x \, \mathcal{L}_0$. We assume that $x$ is a compact spatial direction with an identification $x \to x + 2 \pi R$ and $t$ is a non-compact time direction.

We impose two conditions in order to perform our dimensional reduction:

\begin{enumerate}[label=(\Roman*)]
    \item\label{TT_reduction_one} The seed theory $\mathcal{L}_0$ has no characteristic length scale, and no length scale besides the one associated with the flow parameter $\lambda$ is introduced at any point in our reduction procedure. In particular, this requires that the undeformed theory is a CFT and that we truncate to zero modes along the circle, since excited states are sensitive to the length scale $R$.

    \item\label{TT_reduction_two} We restrict attention to states for which the off-diagonal component of the stress tensor vanishes, $T_{xt} = 0$.
\end{enumerate}

By assumption \ref{TT_reduction_one}, the only relevant energy scale for any process under consideration in the deformed theory $S_\lambda$ is
\begin{align}\label{Lambda_to_lambda}
    \Lambda = \frac{1}{\sqrt{\lambda}} \, .
\end{align}
An infinitesimal change in the characteristic energy $\Lambda$ is equivalent to an infinitesimal scale transformation of the theory, which is generated by the trace of the stress tensor:
\begin{align}
    \frac{d S_\lambda}{d \log ( \Lambda )} = \Lambda \frac{d S_\lambda}{d \Lambda} = \int d^2 x \, \tensor{T}{^\mu_\mu} \, .
\end{align}
We may write $\frac{d S_\lambda}{d \Lambda}$ in another way by using (\ref{Lambda_to_lambda}) and (\ref{TT_flow_eqn_reduction}). Doing this and equating the integrands gives the constraint
\begin{align}\label{TT_trace_flow}
    \tensor{T}{^\mu_\mu} = - \lambda \left( T^{\mu \nu} T_{\mu \nu} - \left( \tensor{T}{^\mu_\mu} \right)^2 \right) \, ,
\end{align}
which is called the trace flow equation in the $\TT$ literature. See, for instance, the incomplete set of references \cite{McGough:2016lol,Kraus:2018xrn,Kraus:2021cwf,Ebert:2022cle,Kraus:2022mnu} where this trace relation plays an important role in holographic analyses of the $\TT$ deformation.

In the coordinates $(x,t)$, we may solve (\ref{TT_trace_flow}) to find
\begin{align}\label{Txx_solved}
    \tensor{T}{^x_x} &= \frac{\tensor{T}{^t_t} + 2 \lambda T^{xt} T_{xt} }{2 \lambda \tensor{T}{^t_t} - 1 } \, \nonumber \\
    &= \frac{\tensor{T}{^t_t}}{2 \lambda \tensor{T}{^t_t} - 1 } \, , 
\end{align}
where in the second step we have used assumption \ref{TT_reduction_two} to drop the dependence on $T_{xt}$.

Using (\ref{Txx_solved}) and $T_{xt} = 0$, the flow equation (\ref{TT_flow_eqn_reduction}) for the action becomes
\begin{align}
    \frac{\partial S_\lambda}{\partial \lambda} = \int dx \, dt \, \frac{ \left( \tensor{T}{^t_t} \right)^2 }{ \frac{1}{2} - \lambda \tensor{T}{^t_t}} \, .
\end{align}
We can now assume that $\tensor{T}{^t_t} \equiv T$ is independent of the $x$ direction, appropriate for a truncation to zero modes as described in \ref{TT_reduction_one}, to obtain a deformation of the $(0+1)$-dimensional quantum mechanics:
\begin{align}\label{QM_TT_reduction}
    \frac{\partial S}{\partial \lambda} = 2 \pi R \int dt \, \frac{T^2}{\frac{1}{2} - \lambda T} \, .
\end{align}
We may absorb the overall volume factor, which arises from performing the integral over $x$, into the normalization of $\lambda$. In the case where $t$ is a Euclidean time direction, this can be written as a flow equation for the Hamiltonian.

Thus we conclude that -- after making the assumptions \ref{TT_reduction_one}, that there is no dimensionful scale in the problem besides $\lambda$, and \ref{TT_reduction_two}, that $T_{xt} = 0$ -- the $\TT$ deformation reduces to the QM-$\TT$ flow equation (\ref{QM_TT_reduction}). The reduction of a $2d$ field theory with a length scale that violates assumption \ref{TT_reduction_one}, such as a generic non-conformal QFT, is more complicated; see for instance Appendix A of \cite{Ebert:2022xfh}.

\subsection{Reduction of Root-$\TT$ Flow}

Unlike the QM-$\TT$ flow, which is a pure $f(H)$ deformation that can be applied to any $(0+1)$-dimensional system with a Hamiltonian, we have only defined the QM-root-$\TT$ flow for systems that possess an $SO(N)$ symmetry in addition to time translation symmetry. Thus in order to obtain such a deformation by dimensional reduction, we will need to restrict the class of theories under consideration, rather than working with a stress tensor for a fairly general theory as we did for the reduction of $\TT$ in the preceding subsection.

Let us therefore focus on two dimensional field theories whose field content is a collection of scalar fields $\phi^i$, $i = 1 , \ldots , N$, which involve only first derivative couplings. This means that the Lagrangian can only be a function of the traces $Y_1$ and $Y_2$ that were defined in equation (\ref{intro_scalar_Xi}). We assume that the $x$ direction is compact with periodicity
\begin{align}
    x \sim x + 2 \pi R \, , 
\end{align}
so that the fields $\phi^i$ admit a Fourier decomposition
\begin{align}\label{fourier_phi}
    \phi^i ( x, t ) = \sum_{n = - \infty}^{\infty} c^i_n ( t ) \exp \left( \frac{i n x}{R} \right) \, ,
\end{align}
subject to the condition $\left( c_n^i \right)^\ast = c_{-n}^i$, if the $\phi^i$ are real.

Let us focus on a single harmonic in the expansion (\ref{fourier_phi}) with mode number $m$. Because of the reality constraint, if we allow $c_m^i$ to be non-zero, we must also let $c_{-m}^i$ be non-zero. Thus we will consider the two modes $\pm m$ and set all other coefficients to zero. If the fields $\phi^i$ have only these two contributions, then the Lorentz invariant $Y_1$ is given by
\begin{align}\label{Y1_one_mode}\hspace{-10pt}
    Y_1 &= \frac{2 m^2}{R^2} c_m^i \cbar_{m, i} - e^{2 i m x / R} \left( \frac{m^2}{R^2} c_m^i c_{m, i} + \dot{c}_m^i \dot{c}_{m, i} \right)  + e^{-2 i m x / R} \left( \frac{m^2}{R^2} \cbar_m^i \cbar_{m, i} + \dot{\cbar}_m^i \dot{\cbar}_{m, i} \right) - 2 \dot{c}_m^i \dot{\cbar}_{m, i} \, .
\end{align}
Integrating (\ref{Y1_one_mode}) over the circle with a convenient choice of normalization, we find
\begin{align}
    - \frac{1}{8 \pi R} \int \, dt \, \int_{0}^{2 \pi R} \, dx \, Y_1 = \frac{1}{2} \int dt \, \left( \dot{c}_m^i \dot{\cbar}_{m, i} - \frac{m^2}{R^2} c_m^i \cbar_{m, i} \right) \, , 
\end{align}
which is the action for a $(2N)$-dimensional harmonic oscillator with unit mass and a frequency which is set by $m$ and $R$. In the notation of Section \ref{sec:general_f_H_J2}, we would write this as
\begin{align}
    - \frac{1}{8 \pi R} \int \, dt \, \int_{0}^{2 \pi R} dx \, Y_1 = \frac{1}{2} \int dt \, \left( X_2 - X_1 \right) \, , 
\end{align}
where $X_2, X_1$ are the $SO(N)$ invariants for the scalars corresponding to the kinetic and potential energies, respectively. This is the familiar statement that, under Kaluza-Klein compactification, the higher harmonics of a field in $(d+1)$ dimensions produce a tower of massive states in $d$ dimensions.

As $Y_2$ is also a singlet under $SO(N)$ rotations of the $N$ scalar fields in two dimensions, one may wonder whether it, too, reduces to an $SO(N)$ invariant of the quantum mechanical theory -- that is, some other function of $X_1, X_2$, and $X_3$. In the case of a single scalar field $\phi$, whose $m$-th Fourier mode is a single complex function of time $c_m$, one finds that
\begin{align}
    - \frac{1}{8 \pi R} \int_{0}^{2 \pi R} dx \, Y_2 = - \frac{3 m^4}{2 R^4} X_1^2 - \frac{3}{2} X_2^2 + \frac{3 m^2}{R^2} X_1 X_2 - \frac{m^2}{2 R^2} X_3^2 \, , 
\end{align}
which is indeed a function of the three $SO(2)$ invariants for a $(0+1)$-dimensional theory. However, if we begin with two or more scalar fields $\phi^i$, which descend to two or more complex $c^i_m ( t )$, then the reduction of $Y_2$ is \emph{not} an $SO(N)$ invariant. It is straightforward to see this by considering the dimensional reduction of one \emph{complex} scalar $\phi$, which has a Fourier expansion (\ref{fourier_phi}) where the modes $c_m$ and $c_{-m}$ are independent complex-valued functions of time. In this case, the dimensionally reduced term $Y_2^{(\text{red})}$ in the $(0+1)$-dimensional action is
\begin{align}\hspace{-40pt}\label{Y2_reduced}
    Y_2^{(\text{red})} &= - \frac{1}{8 \pi R} \int_0^{2 \pi R} dx \, Y_2 \, \nonumber \\
    &= - \frac{m^4}{4 R^4} \left( c_{-m}^2 \cbar_{-m}^2 + 4 c_{-m} \cbar_{-m} c_m \cbar_m + c_m^2 \cbar_m^2 \right) - \frac{1}{4} \left( \dot{c}_{-m}^2 \dot{\cbar}_{-m}^2 + 4 \dot{c}_{-m} \dot{\cbar}_{-m} \dot{c}_m \dot{\cbar}_m + \dot{c}_m^2 \dot{\cbar}_m^2 \right) \nonumber \\
    &\quad - \frac{m^2}{8 R^2} \Bigg[ \cbar_{-m}^2 \dot{c}_{-m}^2 +c_m^2 \dot{\cbar}_m^2 - 2 c_m \left(\cbar_m \dot{c}_{-m} \dot{\cbar}_{-m}+\dot{\cbar}_m \left(2 c_{-m} \dot{\cbar}_{-m}+\cbar_m \dot{c}_m\right)\right) \nonumber \\
    &\quad + 2 \cbar_{-m} \left(\dot{\cbar}_m \left(c_m \dot{c}_{-m}-c_{-m} \dot{c}_m\right)-\dot{c}_{-m} \left(c_{-m} \dot{\cbar}_{-m}+2 \cbar_m \dot{c}_m\right)\right)+\left(c_{-m} \dot{\cbar}_{-m}+\cbar_m \dot{c}_m\right)^2\Bigg] \, .
\end{align}
The dimensionally reduced expression (\ref{Y2_reduced}) is not invariant under the transformation
\begin{align}
    c_m \to c_m + \epsilon c_{-m} \, , \qquad c_{-m} \to c_{-m} - \epsilon c_m \, .
\end{align}
This $U(1)$ rotation which acts on the coefficient functions $c_m$, $c_{-m}$ is part of the $SO(4)$ symmetry group for the four real bosons in the quantum mechanical theory. Since the reduction of $Y_2$ is not invariant under this $U(1)$ action, it is not $SO(4)$ invariant and thus cannot be written as a function of the invariants $X_1, X_2, X_3$.

However, we can impose an additional constraint between the Fourier modes of the complex scalar $\phi$ which will both eliminate some of the degrees of freedom and also restore rotational symmetry after performing the dimensional reduction. To do this, we will simply set one of the two modes $c_m$ and $c_{-m}$ equal to zero, while leaving the other as an unconstrained complex function of time. For concreteness we will choose to set the negatively-moded coefficient to zero:
\begin{align}\label{half_dof_constraint}
    c_{-m} = 0 \, .
\end{align}
Rather than a complex scalar field, if we had instead expanded two real scalar fields $\phi^1$, $\phi^2$ in a trigonometric Fourier series
\begin{align}
    \phi^i ( x, t ) = \sum_{n=0}^{\infty} \left( a_n^i ( t ) \cos \left( \frac{n x}{R} \right) + b_n^i ( t ) \sin \left( \frac{n x}{R} \right) \right) \, , 
\end{align}
and then truncated to a single Fourier mode $n = m$, the corresponding constraint is
\begin{align}
    b_m^1 = - a_m^2 \, , \qquad a_m^1 = b_m^2 \, .
\end{align}
We then see that the effect of the constraint is simply to eliminate half of the degrees of freedom by correlating the cosine and sine terms in the Fourier expansion in a particular way. After imposing this constraint, the invariant $Y_2$ takes the form
\begin{align}\label{Y2_constrained}
    Y_2 = \frac{1}{2 R^4} \left[ -2 m^2 R^2 \cbar_{m} c_m \dot{\cbar}_{m} \dot{c}_m + R^2 \dot{c}_{m}^2 \left( m^2 \cbar_m^2 + 2 R^2 \dot{\cbar}_m^2 \right) + c_{m}^2 \left( 2 m^4 \cbar_m^2 + m^2 R^2 \dot{\cbar}_m^2 \right) \right] \, .
\end{align}
In particular, note that all dependence on the spatial coordinate $x$ has dropped out of the expression (\ref{Y2_constrained}). Not only will the dimensionally reduced quantity $Y_2^{(\text{red})}$ now be invariant under $SO(2)$ rotations of the two real scalars $c_{m}$ and $\cbar_m$, but more is true: the dimensional reduction of any \emph{function} of $Y_2$ now satisfies
\begin{align}
    - \frac{1}{8 \pi R} \int_0^{2 \pi R} dx \, f \left( Y_2 \right) = - \frac{1}{4} f \left( Y_2 \right) \, , 
\end{align}
because the integral over $x$ is trivial. Similarly, after imposing these constraints, we have
\begin{align}
    Y_1 = \frac{m^2}{R^2} c_m \cbar_{m} - \dot{c}_m \dot{\cbar}_{m} \, , 
\end{align}
which is also independent of $x$. Therefore, any two-dimensional Lagrangian
\begin{align}
    \mathcal{L} = f ( Y_1, Y_2 )
\end{align}
descends under dimensional reduction to an $SO(2)$-invariant Lagrangian for a quantum-mechanical theory by performing a trivial integral over the $x$ direction. We emphasize that this integration would \emph{not} be trivial before imposing the constraint (\ref{half_dof_constraint}). For instance, the function appearing in the $2d$ field theory which is the root-$\TT$ deformation of free scalars generically takes the form
\begin{align}
    \sqrt{ 2 Y_2 - Y_1^2 } = \sqrt{ f_1 e^{- 4 i m x / R} + f_2 e^{- 2 i m x / R} + f_3 + \text{c.c} } \, , 
\end{align}
where $f_1, f_2, f_3$ are expressions involving the functions $c_m ( t )$, etc. The dimensional reduction of this term is not trivial, since obviously the integration over $x$ does not commute with the square root operation:
\begin{align}\label{integral_commute}
    &\int_{0}^{2 \pi R} dx \, \sqrt{ f_1 e^{- 4 i m x / R} + f_2 e^{- 2 i m x / R} + f_3 + \text{c.c} } \nonumber \\
    &\quad \neq \sqrt{ \int_{0}^{2 \pi R} dx \, \left( f_1 e^{- 4 i m x / R} + f_2 e^{- 2 i m x / R} + f_3 + \text{c.c} \right) } \, .
\end{align}
However, after imposing the constraint (\ref{half_dof_constraint}), we do not encounter the obstruction (\ref{integral_commute}) to performing the integrals because the $x$-dependent factors drop out.

In particular, this means that we may dimensionally reduce the modified scalar theory of equation (\ref{mod_scalar_theory}) which was obtained from applying the $2d$ root-$\TT$ deformation to a seed theory of two free scalars, finding
\begin{align}
    &- \frac{1}{8 \pi R} \int_0^{2 \pi R} \, dx \, \left( \frac{\cosh ( \gamma )}{2} Y_1 + \frac{\sinh ( \gamma )}{2} \sqrt{ 2 Y_2 - Y_1^2 } \right) \, \nonumber \\
    &\quad \sim \frac{\cosh ( \gamma )}{2} ( X_2 - X_1 ) + \frac{\sinh ( \gamma )}{2} \sqrt{ 4 X_3^2 + \left( X_2 - X_1 \right)^2 } \, ,
\end{align}
which is the ModMax oscillator of equation (\ref{deformed_oscillator_solution}). Therefore, in this limit we have engineered the solution to the QM-root-$\TT$ flow from dimensional reduction of the solution to the ordinary root-$\TT$ flow. It also follows that, in this limit, the operators driving the two flows are related -- the root-$\TT$ operator of the parent field theory descends by dimensional reduction to the combination $\sqrt{ E^2 - J^2 }$ in the quantum mechanical theory.

The preceding argument straightforwardly generalizes to dimensional reduction of $N$ complex scalars $\phi^i$. To summarize, let us reiterate the two assumptions which are needed to obtain the QM-root-$\TT$ deformation by dimensional reduction:

\begin{enumerate}[label = ($\widetilde{\mathrm{\Roman*}}$)]

    \item\label{rtt_assumption_one} We may restrict attention to a single Fourier mode of our scalar fields on the spatial circle. That is, if we expand our $N$ complex scalar fields $\phi^i$ as
    \begin{align}
        \phi^i ( x, t ) = \sum_{n = - \infty}^{\infty} c^i_n ( t ) \exp \left( \frac{i n x}{R} \right) \, ,
    \end{align}
    then we consider only the contribution from the modes $n = \pm m$ for some $m \neq 0$.

    \item\label{rtt_assumption_two} We may eliminate half of the modes by imposing the constraints
    \begin{align}\label{half_dof_constraint_later}
        c_{-m}^i = 0 \, ,
    \end{align}
    for each $i$.
\end{enumerate}
Comparing the two assumptions \ref{rtt_assumption_one} - \ref{rtt_assumption_two} for root-$\TT$ to the preceding assumptions \ref{TT_reduction_one} - \ref{TT_reduction_two} for $\TT$, it is now clear why the QM-$\TT$ deformation does not commute with the QM-root-$\TT$ deformation, despite the fact that their parent flows in two spacetime dimensions do commute. The dimensional reduction of the $\TT$ deformation required that we begin with a CFT and that we do not introduce any characteristic length scale, whereas the reduction of root-$\TT$ explicitly requires that we truncated to an excited state with mode number $m$ which introduces a length scale $\frac{m}{R}$. 

We point out that it would be possible to obtain a two-parameter family of $(0+1)$-dimensional theories $L_{\gamma, \lambda}$ by beginning with the $2d$ theory (\ref{two_parameter_scalar_family}) and dimensionally reducing under the assumptions \ref{rtt_assumption_one} - \ref{rtt_assumption_two} above. However, the resulting family of doubly-deformed theories will not satisfy a QM-$\TT$ flow in $\lambda$.

Finally, it follows from the argument above that the ModMax oscillator can also be obtained from a dimensional reduction of the usual ModMax theory from $(3+1)$-dimensions to $(0+1)$-dimensions. This dimensional reduction proceeds in two steps. First one dimensionally reduces the ModMax theory from $(3+1)$-dimensions to $(1+1)$-dimensions, either by considering a particular limit of plane wave scattering \cite{Conti:2022egv} or by compactifying on a $T^2$ and studying the $2d$ theory of the scalars which descend from Wilson lines on the torus. This first reduction step yields the modified scalar theory. Then one performs the second reduction step described above, by considering the two real scalars as a single complex scalar, Fourier-expanding on a spatial circle, and imposing the constraint (\ref{half_dof_constraint_later}). This yields the $(0+1)$-dimensional ModMax oscillator.

\section{Supersymmetric Flows}\label{sec:SUSY}

Supersymmetry is a powerful tool that offers enhanced analytic control over both quantum field theories and lower-dimensional theories of quantum mechanics; the latter are the primary focus of this work. It is reasonable to expect that any deformation of a supersymmetric $(0+1)$-dimensional system by commuting conserved quantities will preserve supersymmetry. For instance, the QM-root-$\TT$ deformation of the harmonic oscillator to produce the ModMax oscillator can be realized as a deformation of the Hamiltonian:
\begin{align}
    H_0 \longrightarrow H_\gamma = f \left( H_0, J_0^2 \right) \, , 
\end{align}
where $H_0$ is the undeformed Hamiltonian and $J_0$ is the angular momentum. For any such deformation, if the Hamiltonian can be written as
\begin{align}
    H_0 = \{ Q , Q^\dagger \}
\end{align}
for a supercharge $Q$, then since all of $Q, Q^\dagger, H_0$, and $J_0^2$ are commuting, we have $[ H_\gamma, Q ] = 0$ and thus the supercharges will remain conserved quantities in the deformed theory.

However, the additional control provided by supersymmetry is most useful when it is made manifest by a superspace construction. It is therefore desirable to find manifestly supersymmetric presentations of any deformations of interest. In the case of the usual $\TT$ deformation in two spacetime dimensions, this was accomplished in the series of works \cite{Baggio:2018rpv,Chang:2018dge,Jiang:2019hux,Chang:2019kiu,Ferko:2021loo}; see also \cite{Coleman:2019dvf,Lee:2021iut,Lee:2023uxj} for other approaches to $\TT$ in theories with supersymmetry and \cite{Jiang:2019trm} for the extension to $J \overline{T}$ deformations. This superspace formalism has been applied to understand correlation functions in $\TT$-deformed supersymmetric field theories \cite{He:2019ahx,Ebert:2020tuy}, supersymmetric black holes \cite{Manschot:2022lib}, and the $2d$ Volkov-Akulov theory \cite{Cribiori:2019xzp}.

A similar superspace construction can be employed to present manifestly supersymmetric $\TT$-like flows in other spacetime dimensions, such as for $3d$ and $4d$ $\mathcal{N} = 1$ gauge theories of Born-Infeld and ModMax-Born-Infeld type \cite{Ferko:2019oyv,Ferko:2022iru,Ferko:2023sps}. Likewise, the dimensional reduction of the $\TT$ flow for conformal seed theories, which gives the QM-$\TT$ deformation of quantum mechanics, can also be supersymmetrized in the same way \cite{Ebert:2022xfh}.

One would like to extend these superspace constructions of $\TT$ to root-$\TT$. Thus far, this has only been done for the $4d$ root-$\TT$-like deformation of gauge theories, which produces the supersymmetric extension of the ModMax theory \cite{Ferko:2023ruw}. Our goal in this section will be to complete the analogous construction for theories of $\mathcal{N} = 2$ supersymmetric quantum mechanics with an additional $SO(N)$ symmetry.

\subsubsection*{\ul{\it Conventions}}

We work in $\mathcal{N} = 2$ superspace for the $(0+1)$-dimensional spacetime with a single bosonic time coordinate $t$ and a complex Grassmann coordinate $\theta$ whose complex conjugate is written $\overbar{\theta}$. The full set of coordinates for our superspace is written as $(t , \theta , \overbar{\theta} )$, and the action $S$ of a theory is written as a full superspace integral of the form
\begin{align}
    S = \int dt \, d \thetab \, d \theta \, \mathcal{A} \, , 
\end{align}
where $\mathcal{A}$ is a superspace Lagrangian density. The derivatives with respect to the superspace coordinates $(t, \theta, \overbar{\theta})$ will be written as $\partial_t$, $\frac{\partial}{\partial \theta}$, and $\frac{\partial}{\partial \thetab}$, respectively. We combine these derivatives into the natural supercovariant derivative operators $D$ and $\Dbar$ defined by
\begin{align}
    D = \frac{\partial}{\partial \theta} - i \thetab \partial_t \, , \qquad \Dbar = \frac{\partial}{\partial \thetab} - i \theta \partial_t \, .
\end{align}
These supercovariant derivatives satisfy the algebra
\begin{align}\label{susy_algebra}
    \left\{ D , \Dbar \right\} = - 2 i \partial_t \, , 
\end{align}
and the derivatives are nilpotent, so that
\begin{align}
    D^2 = \Dbar^2 = 0 \, .
\end{align}
In addition to the supercovariant derivatives, we also introduce the supercharges
\begin{align}
    Q = \frac{\partial}{\partial \theta} + i \thetab \partial_t \, , \qquad \Qbar = \frac{\partial}{\partial \thetab} + i \theta \partial_t \, .
\end{align}
Throughout this section, we will restrict attention to theories of a collection of real superfields $X^i ( t , \theta , \thetab )$. We take the component expansion of such a superfield to be
\begin{align}
    X^i ( t, \theta , \thetab ) = x^i ( t ) + \theta \psi^i ( t ) - \thetab \psib^i ( t ) + \theta \thetab f^i ( t ) \, .
\end{align}
The field content of the superfield $X^i$, therefore, consists of a real boson $x^i$, a complex fermion $\psi^i$, and an auxiliary field $f^i$. We choose the convention that complex conjugation of a product of Grassmann quantities reverses their order so that, for instance,
\begin{align}
    \left( \theta \psi^i \right)^\ast = \psi^{i \ast} \theta^\ast = \psib \thetab \, , 
\end{align}
and with this convention we see that the superfield $X^i$ satisfies the reality condition
\begin{align}
    \left( X^i \right)^\ast = X^i \, .
\end{align}

\subsection{Construction of Conserved Quantities}

The bosonic flows of Section \ref{sec:bosonic} were built from two conserved quantities: the energy $E$ which is the Noether current associated with time translations, and the angular momentum components $J_{mn}$ which are associated with $SO(N)$ rotations. Our first task will be to construct conserved superfield quantities $\mathcal{E}$ and $\mathcal{J}_{mn}$ which contain the energy $E$ and $J_{mn}$, respectively. To do this, we will use a superfield version of the Noether procedure, following the strategy of \cite{Ebert:2022xfh} (see also the earlier work \cite{Clark:2001zv} for a similar approach).

We restrict attention to actions obtained from a superspace Lagrangian with the following dependence on the scalar superfields $X^i$ and their derivatives:
\begin{align}\label{susy_functional_form}
    S = \int dt \, d \thetab \, d \theta \, \mathcal{A} \left( X^i, D X^i, \Dbar X^i, D \Dbar X^i , \Dbar D X^i \right) \, .
\end{align}
We do not allow the superspace Lagrangian to depend on combinations involving three or more supercovariant derivatives of the $X^i$. However, this functional form (\ref{susy_functional_form}) does allow dependence on the first time derivatives $\partial_t X^i = \dot{X}^i$, since
\begin{align}
    \dot{X}^i = \frac{i}{2} \left( D \Dbar X^i + \Dbar D X^i \right) \, , 
\end{align}
using the $\mathcal{N} = 2$ SUSY algebra (\ref{susy_algebra}). In response to a general variation $\delta X^i$ of the scalar superfields, the change in the superspace Lagrangian $\mathcal{A}$ is given by
\begin{align}\label{general_susy_qm_variation}
    \delta \mathcal{A} &= \delta X^i \, \frac{\delta \mathcal{A}}{\delta X^i}  + \delta ( D X^i ) \, \frac{\delta \mathcal{A}}{\delta ( D X^i ) }  + \delta ( \Dbar X^i ) \, \frac{\delta \mathcal{A}}{\delta ( \Dbar X^i ) } + \delta ( D \Dbar X^i ) \, \frac{\delta \mathcal{A}}{\delta ( D \Dbar X^i ) } \nonumber \\
    &\qquad + \delta ( \Dbar D  X^i ) \, \frac{\delta \mathcal{A}}{\delta ( \Dbar D  X^i ) }  \, .
\end{align}
Using the product rule for supercovariant derivatives, $\delta \mathcal{A}$ can be written as
\begin{align}\label{susy_variation_ibp}
    \hspace{-10pt} &\delta \mathcal{A} = D \left( \delta X^i  \, \frac{\delta \mathcal{A}}{\delta ( D X^i ) } \right) + \Dbar \left( \delta X^i  \, \frac{\delta \mathcal{A}}{\delta ( \Dbar X^i ) } \right) + D \left( \frac{\delta \mathcal{A}}{\delta (D \Dbar X^i)} \Dbar \left( \delta X^i \right) \right)  \nonumber \\
    &+ \Dbar \left( \delta X^i D \left( \frac{\delta \mathcal{A}}{\delta (D \Dbar X^i)} \right) \right) + \Dbar \left( \frac{\delta \mathcal{A}}{\delta  (\Dbar D  X^i)} D \left( \delta X^i \right) \right) + D \left( \delta X^i \Dbar \left( \frac{\delta \mathcal{A}}{\delta (\Dbar D X^i)} \right) \right) \nonumber \\
    &- \delta X^i \left( - \frac{\delta \mathcal{A}}{\delta X^i} + D \left( \frac{\delta \mathcal{A}}{\delta ( D X^i ) } \right) + \Dbar \left( \frac{\delta \mathcal{A}}{\delta ( \Dbar X^i ) } \right) + D \Dbar \left( \frac{\delta \mathcal{A}}{\delta ( D \Dbar X^i ) } \right) + \Dbar D  \left( \frac{\delta \mathcal{A}}{\delta (  \Dbar D  X^i ) } \right) \right) \, .
\end{align}
The last line of (\ref{susy_variation_ibp}) is precisely the expression that one would obtain by demanding that $\delta S = 0$ for the field variation $\delta X^i$ around a given trajectory $X^i$, which is the condition that defines the superspace equation of motion:
\begin{align}\label{susy_qm_eom}
    \frac{\delta \mathcal{A}}{\delta X^i} = D \left( \frac{\delta \mathcal{A}}{\delta ( D X^i ) } \right) + \Dbar \left( \frac{\delta \mathcal{A}}{\delta ( \Dbar X^i ) } \right) + D \Dbar \left( \frac{\delta \mathcal{A}}{\delta ( D \Dbar X^i ) } \right) + \Dbar D  \left( \frac{\delta \mathcal{A}}{\delta (  \Dbar D  X^i ) } \right) \, .
\end{align}
Therefore, when the superfields $X^i$ are on-shell, the variation (\ref{susy_variation_ibp}) is
\begin{align}\label{susy_variation_on_shell}
    \delta \mathcal{A} &\simeq D \left( \delta X^i  \, \frac{\delta \mathcal{A}}{\delta ( D X^i ) } \right) + \Dbar \left( \delta X^i  \, \frac{\delta \mathcal{A}}{\delta ( \Dbar X^i ) } \right) + D \left( \frac{\delta \mathcal{A}}{\delta (D \Dbar X^i)} \Dbar \left( \delta X^i \right) \right)  \nonumber \\
    &+ \Dbar \left( \delta X^i D \left( \frac{\delta \mathcal{A}}{\delta (D \Dbar X^i)} \right) \right) + \Dbar \left( \frac{\delta \mathcal{A}}{\delta  (\Dbar D  X^i)} D \left( \delta X^i \right) \right) + D \left( \delta X^i \Dbar \left( \frac{\delta \mathcal{A}}{\delta (\Dbar D X^i)} \right) \right) \, ,
\end{align}
where we have introduced the symbol $\simeq$ to mean equality up to terms which vanish when the superspace equations of motion are satisfied.

Equation (\ref{susy_variation_on_shell}) will be the starting point for computing the conserved superfields which we need in order to construct our superspace flow. In both cases, we consider the variations $\delta X^i$, $\delta \mathcal{A}$ associated with the generator of a symmetry of the action. For such variations, the on-shell condition (\ref{susy_variation_on_shell}) can be expressed as the conservation equation for an appropriate superfield, exactly as in the ordinary (non-superspace) Noether procedure.

\subsubsection*{\ul{\it Time Translations}}

First we will consider the variation of superfields which is implemented by a shift in the time coordinate,
\begin{align}
    t \to t + \delta t \, , 
\end{align}
under which we have the transformations
\begin{align}\label{time_noether_susy}
    \delta \mathcal{A} &= \left( \delta t \right) \partial_t \mathcal{A} = \frac{i}{2} \left( \delta t \right) \left( D \Dbar \mathcal{A} + \Dbar D \mathcal{A} \right) \, , \nonumber \\
    \delta X^i &= ( \delta t ) \dot{X}^i = \frac{i}{2} \left( \delta t \right) \left( D \Dbar X^i + \Dbar D X^i \right) \, .
\end{align}
Here we have used the algebra of supercovariant derivatives (\ref{susy_algebra}). Substituting these variations into the on-shell condition (\ref{susy_variation_on_shell}) and simplifying, one arrives at a superspace conservation equation of the form
\begin{align}\label{susy_conservation_E}
    D \cEb + \Dbar \cE = 0 \, ,
\end{align}
where
\begin{align}\label{susy_Q_def}
    \cE &= i \left[ ( D \Dbar X^i + \Dbar D X^i ) \left( \frac{\delta \mathcal{A}}{\delta ( \Dbar X^i ) } + D \left( \frac{\delta \mathcal{A}}{\delta ( D \Dbar X^i ) } \right) \right) + \frac{\delta \mathcal{A}}{\delta ( \Dbar D X^i ) } D ( \Dbar D X^i ) - D \mathcal{A} \right] \,  , \nonumber \\
    \cEb &= i \left[ ( D \Dbar X^i + \Dbar D X^i ) \left( \frac{\delta \mathcal{A}}{\delta ( D X^i ) } + \Dbar \left( \frac{\delta \mathcal{A}}{\delta ( \Dbar D X^i ) } \right) \right) + \frac{\delta \mathcal{A}}{\delta ( D \Dbar X^i ) } \Dbar ( D \Dbar X^i ) - \Dbar \mathcal{A} \right] \, ,
\end{align}
as was worked out in \cite{Ebert:2022xfh}. The overall factor of $i$ is a choice of normalization which we will see is natural when we examine the component expansion of these superfields.

\subsubsection*{\ul{\it $SO(N)$ Rotations}}

Following the steps we performed in Section \ref{sec:general_f_H_J2} for bosonic flows, we now make the additional assumption that our supersymmetric theory is also invariant under $SO(N)$ rotations of the superfields $X^i$:
\begin{align}
    X^i ( t , \theta , \thetab ) \longrightarrow \tensor{R}{^i_j} X^j ( t, \theta, \thetab ) \, , \qquad R \in SO ( N ) \, .
\end{align}
As in the bosonic setting, the generators $\delta^{(nm)}$ of these rotations act on the superfields as
\begin{align}
    \delta^{(nm)} X_i = \epsilon_{ij}^{(nm)} X^j \, , 
\end{align}
where $\epsilon_{ij}^{(nm)}$ is given in equation (\ref{epsilon_defn}). We will assume that the superspace Lagrangian is invariant under such a rotation (rather than transforming as a total derivative), so that $\delta^{(nm)} \mathcal{A} = 0$. Then the surviving terms in the on-shell variation (\ref{susy_variation_on_shell}) are
\begin{align}
    0 &\simeq \epsilon_{ij}^{(nm)} \Bigg[ D \left( X^j \, \frac{\delta \mathcal{A}}{\delta ( D X^i ) } \right) + \Dbar \left( X^j  \, \frac{\delta \mathcal{A}}{\delta ( \Dbar X^i ) } \right) + D \left( \frac{\delta \mathcal{A}}{\delta (D \Dbar X^i)} \Dbar \left( X^j \right) \right)  \nonumber \\
    &+ \Dbar \left( X^j D \left( \frac{\delta \mathcal{A}}{\delta (D \Dbar X^i)} \right) \right) + \Dbar \left( \frac{\delta \mathcal{A}}{\delta  (\Dbar D  X^i)} D \left( X^j \right) \right) + D \left(  X^j \Dbar \left( \frac{\delta \mathcal{A}}{\delta (\Dbar D X^i)} \right) \right) \Bigg] \, .
\end{align}
This can be written as a conservation law of the form
\begin{align}
    D \overbar{\mathcal{J}}_{nm} + \Dbar \mathcal{J}_{nm} = 0 \, , 
\end{align}
where
\begin{align}
    \mathcal{J}_{nm} &= \left(  X^n  \, \frac{\delta \mathcal{A}}{\delta ( \Dbar X^m ) } +  \frac{\delta \mathcal{A}}{\delta  (\Dbar D  X^n)} D X^m  + X^n D \left( \frac{\delta \mathcal{A}}{\delta (D \Dbar X^m)} \right)  \right) - \left( n \longleftrightarrow m \right)  \, , \nonumber \\
    \overbar{\mathcal{J}}_{nm} &= \left( X^n \, \frac{\delta \mathcal{A}}{\delta ( D X^m ) } + \frac{\delta \mathcal{A}}{\delta (D \Dbar X^n)} \Dbar  X^m  +  X^n \Dbar \left( \frac{\delta \mathcal{A}}{\delta (\Dbar D X^m)} \right) \right) - \left( n \longleftrightarrow m \right)  \, .
\end{align}

\subsubsection*{\ul{\it Example of Conserved Quantities for Superspace Oscillator}}

To gain intuition for the superfields $\mathcal{E}$, $\mathcal{J}_{nm}$ constructed by the superspace Noether procedure, we will now examine their component expansions for a simple choice of superspace Lagrangian $\mathcal{A}$. We focus on the $\mathcal{N} = 2$ extension of the theory of an $n$-dimensional harmonic oscillator with coordinates $x^i$. The bosonic coordinates are promoted to superfields $X^i$ which are described by the action
\begin{align}
    S = \frac{1}{2} \int dt \, d \thetab \, d \theta \left( D X^i \, \Dbar X^i + X^i X^i \right) \, .
\end{align}
This can be thought of as a member of a more general class of non-linear sigma models,
\begin{align}\label{general_nlsm}
    S = \int dt \, d \thetab \, d \theta \left( \frac{1}{2} g_{ij} ( X ) D X^i \, \Dbar X^j + W ( X ) \right) \, ,
\end{align}
in the special case where the metric $g_{ij}$ is the flat Euclidean metric and the superpotential $W$ is quadratic in the $X^i$. For any non-linear sigma model of the general form (\ref{general_nlsm}), performing the integral over the Grassmann directions $\theta, \thetab$, and integrating out the auxiliary field using its equation of motion, one finds
\begin{align}
    S = \int dt \, \left( \frac{1}{2} g_{ij} \dot{x}^i \dot{x}^j + i g_{ij} \psib^i \nabla_t \psi^j + \frac{1}{4} R_{ijkl} \psib^i \psi^j \psib^k \psi^l - \frac{1}{2} g^{ij} \partial_i W \partial_j W - \psib^i \psi^j \nabla_i \partial_j W \right) \, , 
\end{align}
where $\partial_i = \frac{\partial}{\partial x^i}$, $R_{ijkl}$ is the Riemann curvature tensor associated with the metric $g_{ij}$, and $\nabla_t$, $\nabla_i$ are the time derivative and target space derivative, respectively, made appropriately covariant with respect to $g_{ij}$. In particular, when $W ( X ) = \frac{1}{2} X^i X^i$, $g_{ij} = \delta_{ij}$, and we set the fermions $\psi^i$ to zero, this reduces to
\begin{align}\label{bosonic_oscillator}
    S = \int dt \, \left( \frac{1}{2} \dot{x}^i \dot{x}_i - \frac{1}{2} x^i x_i \right) \, , 
\end{align}
which is the $N$-dimensional harmonic oscillator that we have used as a seed theory for some of the flows in Section \ref{sec:bosonic}.

The conserved energy superfields $\cE$ and $\cEb$ for this model are given by
\begin{align}
    \cE &= - \frac{i}{2} \left[ \left( D \Dbar X^i + \Dbar D X^i \right) D X^i + D \left( D X^i \, \Dbar X^i  + X^i X^i \right) \right] \nonumber \\
    &= - \frac{i}{2} \left( \Dbar D X^i + 2 X^i \right) D X^i  \, , \nonumber \\
    \cEb &= \frac{i}{2} \left[ \left( D \Dbar X^i + \Dbar D X^i \right) \Dbar X^i - \Dbar \left( D X^i \, \Dbar X^i + X^i X^i \right) \right] \nonumber \\
    &= \frac{i}{2} \left( D \Dbar X^i - 2 X^i \right) \Dbar X^i \, .
\end{align}
It is instructive to see where the familiar quantities for the bosonic sector (\ref{bosonic_oscillator}), such as the total energy, sit within these superfields. To this end, we will again set the fermions to zero and replace the auxiliary fields $f^i$ with their equations of motion. For any sigma model (\ref{general_nlsm}) with trivial metric $g_{ij} = \delta_{ij}$, the equation of motion for the auxiliary is
\begin{align}
    f^i = - \frac{\partial W}{\partial x_i} \, , 
\end{align}
which in our case gives $f^i = - x^i$. After making this substitution, the $\theta$-expansions of the energy superfields $\cE, \cEb$ with $\psi^i = 0$ are given by
\begin{align}\label{susy_oscillator_energies}
    \cE &= \frac{i}{2} \thetab \left( x^i x_i + \dot{x}^i \dot{x}_i \right) \, , \nonumber \\
    \cEb &= - \frac{i}{2} \theta \left( x^i x_i + \dot{x}^i \dot{x}_i \right) \, .
\end{align}
Therefore we see that -- when restricting to the on-shell bosonic sector -- the usual total energy $E = \frac{1}{2} \left( \dot{x}^i \dot{x}_i + x^i x_i \right)$ sits in the middle components of the superfields $\cE$ and $\cEb$. The overall factors of $i$ have been included to ensure that $\left( \cE \right)^\ast = \cEb$, which means that these superfields are real. In particular, note that
\begin{align}\label{EE_int}
    \int d \thetab \, d \theta \, \cEb \cE = \frac{1}{4} \left( \dot{x}^i \dot{x}_i + x^i x_i \right)^2 = E^2 \, .
\end{align}
Likewise, the angular momentum superfields for this superspace Lagrangian are
\begin{align}\label{susy_oscillator_Js}
    \mathcal{J}_{nm} &= \frac{1}{2} \left( X^n D X^m - X^m D X^n \right) \, , \nonumber \\
    \overbar{\mathcal{J}}_{nm} &= - \frac{1}{2} \left( X^n \Dbar X^m - X^m \Dbar X^n \right) \, .
\end{align}
With the fermions set to zero and auxiliaries on-shell, these have the expansions
\begin{align}\label{JJ_int}
    \mathcal{J}_{nm} &= - \frac{i}{2} \thetab \left( x_n \dot{x}_m - x_m \dot{x}_n \right) \, , \nonumber \\
    \overbar{\mathcal{J}}_{nm} &= \frac{i}{2} \theta \left( x_n \dot{x}_m - x_m \dot{x}_n \right) \, .
\end{align}
Again, we see that the usual angular momentum $J_{nm}$ sits in the middle components, and that these superfields come along with an overall factor of $i$ to make them real. The top component of the product is
\begin{align}
    \int d \thetab \, d \theta \, \overbar{\mathcal{J}}_{nm} \mathcal{J}^{nm} = \frac{1}{4} \left( x_n \dot{x}_m - x_m \dot{x}_n \right) \left( x^n \dot{x}^m - x^m \dot{x}^n \right) = J^2 \, .
\end{align}

\subsection{Superspace Root-$\TT$-Like Deformation}

Having constructed superfields $\mathcal{E}$, $\mathcal{J}_{nm}$ which extend the usual energy $E$ and angular momentum components $J_{nm}$, one could now proceed as in Section \ref{sec:general_f_H_J2} and consider deformations of the superspace Lagrangian by arbitrary functions of these quantities:
\begin{align}\label{susy_f_E_Jsq_deformations}
    \frac{\partial \mathcal{A}}{\partial \lambda} = f \left( \cEb , \mathcal{E}, \overbar{\mathcal{J}}^{nm} , \mathcal{J}_{nm}  \right) \, .
\end{align}
In the special case where the function $f$ depends only on energies but not on angular momenta, this class of flows is related to the supersymmetric versions of $f(H)$ deformations. For instance, the flow equation which was proposed in \cite{Ebert:2022xfh},
\begin{align}
    \frac{\partial \mathcal{A}}{\partial \lambda} = \frac{ \cE \cEb}{\frac{1}{2} - 2 \lambda \Dbar \cE} \, , 
\end{align}
yields a manifestly supersymmetric extension of the QM-$\TT$ deformation.

Within the general class of superspace deformations (\ref{susy_f_E_Jsq_deformations}), we would like to single out a flow which gives a supersymmetric version of the QM-root-$\TT$ deformation. More precisely, we would like to find a function $f$ so that the flow (\ref{susy_f_E_Jsq_deformations}) for a theory of $N$ superfields $X^i$ reproduces the flow equation (\ref{squared_root_TT_flow}) for the bosonic components $x^i$ after setting all fermions to zero, eliminating the auxiliary fields using their equations of motion, and integrating out the superspace directions.

One might guess that an appropriate form for such a superfield $f$ would be
\begin{align}\label{guess_susy_root_TT}
    f \overset{?}{=} \sqrt{ \cEb \cE  - \overbar{\mathcal{J}}^{nm} \mathcal{J}_{nm}  } \, , 
\end{align}
since this is the result of promoting each object in the operator $R = \sqrt{ E^2 - J^2 }$ of equation (\ref{rtt_QM_first_mention}) to the corresponding superfield. However, this guess fails. Although the top component of the superfield $\cEb \cE - \overbar{\mathcal{J}}^{nm} \mathcal{J}_{nm} $ indeed matches the expression $E^2 - J^2$ when truncating to the bosons, the top component of the square root of a superfield is not the same as the square root of the top component of a superfield. To find the former, one must evaluate the square root via the finite Taylor series expansion; for instance,
\begin{align}
    \sqrt{ x + \theta \psi } = \sqrt{x} + \frac{1}{2 \sqrt{x}} \theta \psi \, .
\end{align}
By performing this expansion, one finds that the guess (\ref{guess_susy_root_TT}) does not have the correct highest component. In fact, it is not even well-defined, because the lowest components of $\cEb \cE$ and $\overbar{\mathcal{J}}^{nm} \mathcal{J}_{nm}$ actually vanish when $\psi^i = 0$ and auxiliaries are put on-shell.

A better guess is obtained by beginning with the combination $ \cEb  \cE - \overbar{\mathcal{J}}^{nm} \mathcal{J}_{nm} $, whose top component is roughly $E^2 - J^2$, and divide by a superfield whose \emph{lowest} component is $\sqrt{E^2 - J^2}$. One way to achieve this is by defining
\begin{align}\begin{split}\label{susy_root_TT_later}
    &\quad f \left( \cEb , \mathcal{E}, \overbar{\mathcal{J}}^{nm} , \mathcal{J}_{nm}  \right) = \mathcal{R} \, , \\
    \mathcal{R} \equiv &\frac{\cEb  \cE - \overbar{\mathcal{J}}^{nm} \mathcal{J}_{nm}}{ \sqrt{ \left( D \cEb \right) \left( \Dbar \cE \right) - \left( D \overbar{\mathcal{J}}_{nm} \right) \left( \Dbar \mathcal{J}^{nm} \right) } } \, .
\end{split}\end{align}
Equation (\ref{susy_root_TT_later}) defines the supersymmetric QM-root-$\TT$ operator $\mathcal{R}$ that will drive the flows which we are interested in. We note that the functional form of this operator is nearly identical to that of the root-$\TT$-like operator constructed in \cite{Ferko:2023ruw} for $4d$ Abelian gauge theories with $\mathcal{N} = 1$ supersymmetry, which we reproduce here for comparison:
\begin{equation}
    \mathcal{R}^{(4d)} = \frac{\mathcal{J}^{\alpha\dot\alpha}\mathcal{J}_{\alpha\dot\alpha} - \overbar{\mathcal{X}}\mathcal{X}}{\sqrt{([D^{(\gamma},\overbar{D}_{(\dot{\gamma}}]\mathcal{J}^{\delta)}{}_{\dot{\delta})})    [D_{(\gamma},\overbar{D}^{(\dot{\gamma}}]\mathcal{J}_{\delta)}{}^{\dot{\delta})}}} \, .
    \label{susyrootttbar_4d}
\end{equation}
In (\ref{susyrootttbar_4d}), $\mathcal{J}_{\alpha \dot{\alpha}}$ and $\chi$ are the fields in the $4d$ Ferrara-Zumino multiplet \cite{Ferrara:1975}. The role of the Lorentz scalars $\mathcal{J}^{\alpha\dot\alpha} \mathcal{J}_{\alpha\dot\alpha}$ and $\overbar{\mathcal{X}}\mathcal{X}$ in our $(0+1)$-dimensional superspace deformation are played by $\cEb  \cE$ and  $\overbar{\mathcal{J}}^{nm} \mathcal{J}_{nm}$, respectively.

It is straightforward to develop intuition for the operator $\mathcal{R}$ by examining its component expansion, with the fermions set to zero and auxiliary fields put on-shell, using our results above. Using equations (\ref{susy_oscillator_energies}) and (\ref{susy_oscillator_Js}), the bosonic part of this operator is
\begin{align}
    \mathcal{R} = \sqrt{ E^2 - J^2 } \, d \theta \, d \thetab \, , 
\end{align}
where $E$ and $J^2$ are the usual energy and angular momentum for the scalars $x^i$, written for instance in equations (\ref{EE_int}) and (\ref{JJ_int}). It follows that, to leading order, the flow driven by the superspace operator $\mathcal{R}$ reproduces the QM-root-$\TT$ flow for the bosons when the fields $f^i$ satisfy their equations of motion.

Clearly the same conclusion will hold for any superspace model for which
\begin{align}
    \mathcal{E} = i \thetab E \, , \qquad \mathcal{J}_{nm} = - i \thetab J_{nm} \, , 
\end{align}
and likewise for the conjugate quantities $\cEb$, $\overbar{\mathcal{J}}_{nm}$, when the auxiliaries are on-shell. The advantage of the superfield formulation of this flow is that, besides manifestly reproducing the QM-root-$\TT$ for the bosons, we also obtain the deformed theory for the fermions in a way which manifestly preserves off-shell supersymmetry.

\subsection{Manifestly Supersymmetric ModMax oscillator}

We now propose a supersymmetric extension of the ModMax oscillator which is obtained from the QM-root-$\TT$ flow applied to the seed theory of an $n$-dimensional harmonic oscillator. Using the results of the preceding subsection, let the undeformed superspace Lagrangian be
\begin{align}
    \mathcal{A}_0 = \frac{1}{2} \left( D X^i \, \Dbar X^i + X^i X^i \right) \, , 
\end{align}
and define the superfield energies and angular momenta associated with this theory as
\begin{align}
    \cE_0 &= - \frac{i}{2} \left( \Dbar D X^i + 2 X^i \right) D X^i  \, , \nonumber \\
    \cEb_0 &= \frac{i}{2} \left( D \Dbar X^i - 2 X^i \right) \Dbar X^i \, , \nonumber \\
    \mathcal{J}_{nm} &= \frac{1}{2} \left( X^n D X^m - X^m D X^n \right) \, , \nonumber \\
    \overbar{\mathcal{J}}_{nm} &= - \frac{1}{2} \left( X^n \Dbar X^m - X^m \Dbar X^n \right) \, .
\end{align}
We then propose the following action for the supersymmetric ModMax oscillator:
\begin{align}\label{susy_deformed_oscillator}
    S_\gamma = \int dt \, d \thetab \, d \theta \, \left( \cosh ( \gamma ) \mathcal{A}_0 + \sinh ( \gamma ) \frac{\cEb_0  \cE_0 - \overbar{\mathcal{J}}^{nm}_0 \mathcal{J}_{0, nm}}{ \sqrt{ \left( D \cEb_0 \right) \left( \Dbar \cE_0 \right) - \left( D \overbar{\mathcal{J}}_{0,nm} \right) \left( \Dbar \mathcal{J}^{nm}_0 \right) } } \right) \, .
\end{align}
When all fermions are set to zero and the auxiliary field is replaced using its equation of motion, the superspace action (\ref{susy_deformed_oscillator}) reproduces the theory of the ModMax oscillator for the bosons $x^i$.

By construction, to leading order in the flow parameter $\gamma$, the action $S_\gamma$ yields a solution to the flow equation
\begin{align}\label{on_shell_flow}
    \partial_\gamma S_\gamma = \int dt \, d \theta \, d \thetab \, \mathcal{R}_\gamma \, , 
\end{align}
where the operator $\mathcal{R}_\gamma$ is constructed from $S_\gamma$ according to the definition (\ref{susy_root_TT_later}). We expect that this manifestly supersymmetric Modmax oscillator satisfies the flow equation on-shell to all orders in $\gamma$. More precisely, by ``on-shell'' we mean that the equation (\ref{on_shell_flow}) is satisfied if one imposes the equation of motion for the auxiliary field, which is
\begin{align}\label{auxiliary_eom_along_flow}
    f^i = - x^i \, .
\end{align}
In particular, the equation of motion for the auxiliary remains (\ref{auxiliary_eom_along_flow}) at all points along the flow; it does not become deformed with $\gamma$.

It would be interesting to find a direct superspace proof of this all-orders conjecture, which would likely require identifying an appropriate superfield identity which holds when the equation of motion for the auxiliary field is satisfied. Such identities have been used to prove analogous superspace flow equations for the case of manifestly supersymmetric $\TT$ flows \cite{Chang:2019kiu,Ferko:2019oyv}.

\subsection{Comments on $\mathcal{N} = 1$ Supersymmetry}

The reader may wonder why we have chosen to focus on theories with $\mathcal{N} = 2$ supersymmetry. At first glance, it might seem that theories with $\mathcal{N} = 1$ supersymmetry would be a simpler setting in which to understand a manifestly supersymmetric version of the QM-root-$\TT$ deformation, since an $\mathcal{N} = 1$ superfield $X^i$ contains fewer component fields. In particular such a superfield has no auxiliary field $f^i$ in its expansion, which would avoid some of the complications that appeared in the $\mathcal{N} = 2$ analysis.

To conclude this section, let us briefly comment on the reason we are less interested in the $\mathcal{N} = 1$ case. Recall that an $\mathcal{N} = 1$ superspace has a single anticommuting coordinate $\theta$. The expansion of an $\mathcal{N} = 1$ scalar superfield $X^i$ takes the form
\begin{align}
    X^i = x^i + i \theta \psi^i \, .
\end{align}
The supercovariant derivative associated with $\theta$ is
\begin{align}
    D = \frac{\partial}{\partial \theta} - i \theta \frac{\partial}{\partial t} \, ,
\end{align}
which satisfies the algebra
\begin{align}
    \{ D , D \} = - 2 i \partial_t \, .
\end{align}
An integral over $\mathcal{N} = 1$ superspace involves a measure with only a single Grassmann object $d \theta$; for instance, a theory of scalar fields $X^i$ is defined by a superspace integral
\begin{align}
    S = \int dt \, d \theta \, \mathcal{A} \left( X^i , DX^i , \dot{X}^i \right) \, .
\end{align}
Because the action $S$ is commuting and $d \theta$ is anticommuting, the superspace Lagrangian $\mathcal{A}$ must also be anticommuting. As a simple example, the free superspace Lagrangian for such a collection of $\mathcal{N} = 1$ superfields is written
\begin{align}\label{example_n_equals_one_free}
    \mathcal{A} = \frac{i}{2} \dot{X}^i D X^i \, ,
\end{align}
which after performing the integral over $\theta$ yields
\begin{align}
    L = \int d \theta \,  \frac{i}{2} \dot{X}^i D X^i = \frac{1}{2} \left( \dot{x}^i \dot{x}_i + i \psi^i \dot{\psi}^i \right) \, .
\end{align}
The $\mathcal{N} = 1$ superspace Noether procedure yields less elegant formulas than the $\mathcal{N} = 2$ version. Because there is only a single covariant derivative, we cannot obtain a pair of quantities like $\cE$ and $\cEb$, as in the $\mathcal{N} = 2$ case, which satisfy a conservation equation
\begin{align}
    D \cEb + \Dbar \cE = 0 \, .
\end{align}
The only two derivatives for $\mathcal{N} = 1$ are the supercovariant derivative $D$ and the ordinary time derivative $\partial_t$, so the only possible conservation equation involving two terms is
\begin{align}\label{Neq1_cons}
    D \mathcal{E}_\theta + \partial_t \mathcal{E}_t = 0 \, .
\end{align}
Nonetheless, expressions which satisfy this conservation equations can be found using a Noether procedure. The details of this calculation are described in \cite{Ebert:2022xfh}, so we merely quote the results. A pair of charges $\mathcal{E}_t, \mathcal{E}_\theta$ satisfying (\ref{Neq1_cons}) are given by
\begin{align}
    \mathcal{E}_t &= ( D X^i ) \frac{\delta \mathcal{A}}{\delta ( D X^i )} + i ( D X^i ) D \left( \frac{\delta \mathcal{A}}{\delta \dot{X}^i} \right) \, , \nonumber \\
    \mathcal{E}_\theta &= i  \dot{X}^i \left( \frac{\delta \mathcal{A}}{\delta ( D X^i ) } \right) \, .
\end{align}
For the free theory (\ref{example_n_equals_one_free}), these charges are
\begin{align}
    \mathcal{E}_t = i \left( D X^i \right) \dot{X}^i \, , \qquad \mathcal{E}_\theta = - \frac{1}{2} \dot{X}^i \dot{X}^i \, .
\end{align}
A similar calculation to find the Noether charge associated with $SO(N)$ rotation produces the conserved quantities
\begin{align}
    \mathcal{J}_\theta^{nm} = X^n \frac{\delta \mathcal{A}}{\delta ( DX^m )} -  X^m \frac{\delta \mathcal{A}}{\delta ( DX^n )} \, , \qquad \mathcal{J}_t^{nm} = X^n \frac{\delta \mathcal{A}}{\delta \dot{X}^m} - X^m \frac{\delta \mathcal{A}}{\delta \dot{X}^n} \, , 
\end{align}
which satisfy
\begin{align}
    D \left( \mathcal{J}_\theta^{nm} \right) + \partial_t \left( \mathcal{J}_t^{nm} \right) = 0 \, .
\end{align}
For the free theory (\ref{example_n_equals_one_free}), these are
\begin{align}
    \mathcal{J}_\theta^{nm} = \frac{i}{2} \left( X^n \dot{X}^m -  X^m \dot{X}^n \right) \, , \qquad \mathcal{J}_t^{nm} = \frac{i}{2} \left(  X^n D X^m - X^m D X^n \right) \, .
\end{align}
So $\mathcal{J}_\theta^{nm}$ has the ordinary angular momentum in its lowest component and $\mathcal{J}_t^{nm}$ has the ordinary angular momentum in its highest component, up to constant factors.

One could then proceed to construct a guess for a superspace root-$\TT$ operator as we did in the $\mathcal{N} = 2$ case by examining where the combinations $E^2$ and $J^2$ sit within various combinations of superfields. For instance, after setting the fermions to zero, one finds
\begin{align}
    \int d \theta \, \mathcal{E}_t \mathcal{E}_\theta = - \frac{1}{2} \left( \dot{x}^i \dot{x}_i \right)^2 = - 2 E^2 \, , 
\end{align}
where $E = \frac{1}{2} \dot{x}^i \dot{x}_i$ is the energy for the bosons. Likewise,
\begin{align}
    \int d \theta \, \mathcal{J}_\theta^{nm} \mathcal{J}_{t, nm} &= \frac{i}{4} \left( x^n \dot{x}^m - x^m \dot{x}^n \right) \left( x_n \dot{x}_m - x_m \dot{x}_n \right) \nonumber \\
    &= \frac{i}{4} J^2 \, ,
\end{align}
where $J^2$ is the total squared angular momentum for the bosons. The combinations which have the corresponding quantities in their lowest components are
\begin{align}
    \frac{1}{4}  \left[ ( D \mathcal{E}_t ) ( D \mathcal{E}_t ) \right]_{\theta = \psi = 0} &= E^2 \, , \nonumber \\
    4 \left[ ( D \mathcal{J}_t^{nm} ) ( D \mathcal{J}_{t, nm} ) \right]_{\theta = \psi = 0} &= J^2 \, .
\end{align}
The most straightforward proposal for an $\mathcal{N} = 1$ root-$\TT$ operator is therefore
\begin{align}\label{Neq1_rTT}
    \mathcal{R} = \frac{ - \frac{1}{2} \mathcal{E}_t \mathcal{E}_\theta + 4 i \mathcal{J}_\theta^{nm} \mathcal{J}_{t, nm} }{\sqrt{ \frac{1}{4}  ( D \mathcal{E}_t ) ( D \mathcal{E}_t ) - 4 ( D \mathcal{J}_t^{ij} ) ( D \mathcal{J}_{t, ij} )  } } \, .
\end{align}
The combination (\ref{Neq1_rTT}) has the property, by construction, that truncating to the bosons and performing the superspace integral reproduces the usual root-$\TT$ operator $\sqrt{E^2 - J^2}$ for the free theory (\ref{example_n_equals_one_free}).

In principle, this allows us to define and study root-$\TT$-like flows for any $SO(N$) symmetric $(0+1)$-dimensional theory with $\mathcal{N} = 1$ supersymmetry. However, we now run into an issue, because the main case of our interest is supersymmetric extensions of the theory of $N$ bosons $x^i$ subject to a harmonic oscillator potential. In $\mathcal{N} = 1$ superspace, it is rather awkward to write down such a potential for bosons; we cannot simply add a term $X^i X^i$, as we did in the $\mathcal{N} = 2$ case, since the superspace Lagrangian must be a fermion. The only fermionic objects at our disposal are $D$ and $\theta$, and any term in the superspace Lagrangian involving $D X^i$ will introduce time derivatives of the $x^i$. Therefore, the only way to introduce a potential is to consider a superspace Lagrangian of the form
\begin{align}\label{Neq1_harmonic_oscillator}
    \mathcal{A} = \frac{i}{2} \dot{X}^i D X^i + \frac{1}{2} \theta X^i X^i \, .
\end{align}
This is somewhat unsatisfying, since we have resorted to allowing the $\mathcal{A}$ to have explicit dependence on the superspace coordinate $\theta$. This means that the superspace Lagrangian is no longer invariant under superspace translations in the $\theta$ direction. It is, of course, still possible to define a conserved energy and angular momentum for this theory, but we are now essentially studying the ``hybrid'' model
\begin{align}
    S = \left ( \int \, d t \, d \theta \, \frac{i}{2} \dot{X}^i D X^i  \right) + \frac{1}{2} \int dt \, x^i x^i \, , 
\end{align}
since the second term of (\ref{Neq1_harmonic_oscillator}) simply picks out the lowest component of $X^i X^i$.

Because of the unwieldiness of including a potential in $\mathcal{N} = 1$ language, we have chosen not to focus on this case. Instead we primarily study the case of $\mathcal{N} = 2$ theories, where both the kinetic and potential terms can be more comfortably encoded in natural superspace structures.

\section{Conclusion}\label{sec:conclusion}

In this work, we have established several new properties of root-$\TT$-like deformations in quantum mechanical systems. Among our main results are that such deformations can be defined in \emph{any} theory in $(0+1)$ spacetime dimensions with $SO(N)$ symmetry; that the QM-root-$\TT$ flow descends from a particular dimensional reduction of the $2d$ root-$\TT$ flow; and that these deformations can be made manifestly supersymmetric by writing them in $\mathcal{N} = 2$ superspace.

There remain several interesting directions for future research. Perhaps the most obvious is to investigate the quantization of the ModMax oscillator, which will be studied in a separate work. Another obvious direction is to construct a form of the $(0+1)$-dimensional root-$\TT$ operator, or its higher dimensional analogues, with a larger amount of supersymmetry. Because the $4d$ version of the root-$\TT$ flow can be written in $\mathcal{N} = 1$ superspace \cite{Ferko:2023ruw}, by dimensional reduction it is natural to expect that one could formulate supersymmetric root-$\TT$ flows for $2d$ theories with $\mathcal{N} = (2, 2)$ supersymmetry, or for quantum mechanical theories with $\mathcal{N} = 4$. This might facilitate a study of other observables. Although it seems that the most obvious quantity to compute, the Witten index, will not flow under root-$\TT$ -- since all ground states of the undeformed theory remain ground states in the deformed theory with energies which are simply rescaled -- it would interesting to look for other index-like quantities that do flow. 

There is another question which concerns conformal invariance. The usual $2d$ root-$\TT$ operator is classically marginal and therefore preserves classical conformal symmetry when applied to a CFT seed. It is not known whether conformal invariance of the deformed theory persists after quantization. One might investigate an analogue of this question in $(0+1)$-dimensions by considering root-$\TT$-like deformations of conformal (or superconformal) quantum mechanics; see \cite{Britto-Pacumio:1999dnb} for a review of these theories.

Besides these immediate next steps, we outline a few other interesting directions below.

\subsubsection*{\ul{\it Chaos and Deformations of SYK}}

Another interesting direction is to investigate the fate of quantum chaos in theories which are deformed by the $(0+1)$-dimensional version of root-$\TT$. A similar analysis for the quantum mechanical analogue of $\TT$ was carried out in \cite{He:2022ryk}, focusing on the spectral form factor, out-of-time-order correlator, and Krylov complexity. There it was found that, in a certain sense, the chaotic behavior of SYK-like models remains unchanged by the $(0+1)$-dimensional version of $\TT$, and the effect of the deformation is essentially a rescaling of the time parameter. 

It is natural to ask whether something similar is true for the $(0+1)$-dimensional version of root-$\TT$, or if the non-analyticity in this deformation leads to a more dramatic effect on chaos. An immediate issue is that, na\"ively, it does not seem possible to apply our root-$\TT$-like deformation to a model with only fermionic degrees of freedom (like the conventional SYK theory \cite{kitaev_talk,PhysRevLett.70.3339}) since the square root of a Grassmann-valued quantity is not well-defined. However, as pointed out in \cite{Ferko:2022cix}, one can circumvent this issue by considering a model with both bosons and fermions (such as a supersymmetric theory). In this case, the square root appearing in the deformed Lagrangian admits a finite Taylor series expansion in the fermionic quantities.

In particular, one could apply our manifestly supersymmetric root-$\TT$-like deformation to one of the supersymmetric extensions of the SYK theory \cite{Fu:2016vas,Murugan:2017eto,Peng:2017spg}. For instance, one can study the presentation of supersymmetric SYM written in $\mathcal{N} = 2$ superspace \cite{Bulycheva:2018qcp} where the degrees of freedom are a collection of chiral superfields $\Psi_i$ with expansion
\begin{align}
    \Psi_i = \psi_i \left( \tau + \theta \overbar{\theta} \right) + \theta b_i \, , 
\end{align}
where $\psi_i$ are complex fermions and $b_i$ are complex scalars. The superspace Lagrangian for this theory consists of a kinetic F-term of the form $\overbar{\Psi}_i D \Psi_i$ as well as a holomorphic superpotential with random Gaussian couplings. In the undeformed model, the complex scalars $b_i$ are non-dynamical, but they will appear in the expressions for the superfield energy $\mathcal{E}$ and angular momenta $\mathcal{J}_{nm}$. It would be interesting to understand the effect of our $\mathcal{N} = 2$ version of the quantum mechanical root-$\TT$ deformation on this model and what effect (if any) this flow has on observables related to chaos.

\subsubsection*{\ul{\it Holographic Interpretation}}

There has been much progress on understanding stress tensor deformations of $2d$ field theories, and their dimensional reductions to quantum mechanics, in terms of modified boundary conditions in a bulk holographic dual. For instance, applying the usual $2d$ $\TT$ deformation to a CFT seed theory can be interpreted as imposing certain mixed boundary conditions for the metric in the dual $\mathrm{AdS}_3$ spacetime \cite{Guica:2019nzm}.

This can also be understood using the Chern-Simons formulation of $\mathrm{AdS}_3$ gravity; from this perspective, the boundary deformation also defines a new variational principle in which a linear combination of a source and its dual expectation value is held fixed at infinity \cite{Llabres:2019jtx}. These modified Chern-Simons boundary conditions can be used to compute observables in the deformed theory, such as Wilson lines \cite{Ebert:2022ehb}.

Just as $\mathrm{AdS}_3$ gravity can be studied in either the metric formalism or the Chern-Simons formalism, its dimensional reduction to two dimensions can be thought of as either a JT gravity theory or a BF gauge theory. The holographic dual to this $2d$ bulk theory, in either formulation, is a $(0+1)$-dimensional quantum mechanical model. One can therefore apply the $(0+1)$-dimensional version of $\TT$ to this theory of quantum mechanics and find the deformed boundary conditions corresponding to this deformation in either presentation of the bulk dual \cite{Ebert:2022ehb}.

It would be worthwhile to repeat this analysis to find the appropriate modified boundary conditions for JT gravity or BF gauge theory for the quantum mechanical root-$\TT$ deformation. One way to do this is to begin with the root-$\TT$ deformed boundary conditions for $\mathrm{AdS}_3$ gravity, which have been written down in both metric and Chern-Simons variables \cite{Ebert:2023tih}, and perform a circle reduction. However, we have seen that the interesting $(0+1)$-dimensional root-$\TT$ arises from dimensionally reducing an \emph{excited} state on the spatial circle, rather than the more conventional type of dimensional reduction which truncates to the zero mode sector. It would be interesting to understand such a reduction, which produces a massive boundary theory, more deeply from the bulk perspective.

As a final comment, we note that all of this discussion has been motivated by the so-called ``double-trace'' $\TT$ operator, and its marginal version which is the usual root-$\TT$ deformation. However, there is also a ``single-trace'' version of the $\TT$ operator which has been proposed and studied in \cite{Giveon:2017nie,Giveon:2017myj}. Although much of the progress that has been made in understanding this single-trace deformation comes from worldsheet techniques, some observables such as the deformed mass formula can be seen directly from a gravity analysis \cite{Chang:2023kkq}. It would be very exciting if one could find a marginal version of this deformation which could be interpreted as a single-trace version of the root-$\TT$ deformation. We hope to return to this, and the other questions raised above, in future work.

\section*{Acknowledgements}

We thank Stephen Ebert, Mukund Rangamani, Savdeep Sethi, Zhengdi Sun, and Gabriele Tartaglino-Mazzucchelli for helpful discussions. C. F. acknowledges productive conversations on related topics with the participants of the \href{https://sites.google.com/view/intropea2023/home#h.ksnkkrrvun8q}{``Integrability in Low-Supersymmetry Theories''} mini-workshop, as well as the conference \href{https://indico.phys.ethz.ch/event/49/}{``Integrability in Gauge and String Theory 2023''}. C.F. is also grateful to INFN and the University of Padua for hospitality during a visit when part of this work was completed. C. F. is supported by U.S. Department of Energy grant DE-SC0009999 and by funds from the University of California.

\appendix

\section{Non-Commutativity of QM-$\TT$ and QM-Root-$\TT$ Flows}\label{app:other}

In this Appendix, we will show that a model which is obtained by performing the $(0+1)$-dimensional analogues of both the $\TT$ and root-$\TT$ flows depends on the order in which the deformations are applied, unlike the $2d$ version of these flows. That is, the diagram
\begin{center}
\begin{tikzcd}
    {L_0, H_0} && {L_\gamma, H_\gamma} \\
    \\
    {L_\lambda, H_\lambda} & {} & { L_{\lambda, \gamma} , H_{\lambda, \gamma} }
    \arrow["{{\text{QM-Root-} \TT}}", from=1-1, to=1-3]
    \arrow["{{\text{QM-}\TT}}"', from=1-1, to=3-1]
    \arrow["{{\text{QM-Root-} \TT}}"', from=3-1, to=3-3]
    \arrow["{{\text{QM-}\TT}}", from=1-3, to=3-3]
\end{tikzcd}
\end{center}
does not commute.

In this diagram, we have included both the Lagrangian $L$ and Hamiltonian $H$ for each theory. This is because the QM-$\TT$ flow is most commonly defined in terms of the Hamiltonian, whereas we have written the QM-root-$\TT$ deformation in a Lagrangian formulation. Therefore we will need to perform a Legendre transformation at the final step of the sequential flow processes in order to compare the results.

First we recall the definition of the QM-$\TT$ flow \cite{Gross:2019ach,Gross:2019uxi}. Under the $(0+1)$-dimensional $\TT$ deformation, the Hamiltonian obeys the flow equation
\begin{align}\label{QM_TT_Hamiltonian}
    \frac{\partial H}{\partial \lambda} = \frac{H^2}{\frac{1}{2} - 2 \lambda H} \, , 
\end{align}
with the solution
\begin{align}\label{QM_TT_deformed_hamiltonian}
    H ( \lambda ) = \frac{1}{4 \lambda} \left( 1 - \sqrt{ 1 - 8 \lambda H ( 0 ) } \right) \, .
\end{align}
On the other hand, our definition of the QM-root-$\TT$ flow is
\begin{align}
    \frac{\partial L}{\partial \gamma} = \sqrt{ E^2 - J^2 } \, ,
\end{align}
where $E$ and $J$ are the conserved energy and angular momentum, respecitvely, associated with the Lagrangian $L$.

To show that these flows do not commute in general, it suffices to exhibit one counterexample. We will study the combined flows beginning from the harmonic oscillator,
\begin{align}
    L_0 = \frac{1}{2} \left( \dot{x}^i \dot{x}_i - x^i x_i \right) \, , \qquad H_0 = \frac{1}{2} \left( p^i p_i + x^i x_i \right) \, .
\end{align}
Let us write $L_{\gamma, \lambda}$, $H_{\gamma, \lambda}$ for the quantities obtained by first performing the QM-root-$\TT$ deforming and then the $\TT$ deformation, which corresponds to the top-right path through the diagram above. Likewise we write $L_{\lambda, \gamma}$ and $H_{\lambda, \gamma}$ for the Lagrangian and Hamiltonian obtained by taking the bottom-left path through the diagram, first flowing by QM-$\TT$ and then by QM-root-$\TT$.

To follow the top-right path, we first recall that the solution to the root-$\TT$ flow with initial condition $L_0$ is the ModMax oscillator whose Lagrangian is (\ref{deformed_oscillator_solution}) and whose Hamiltonian is
\begin{align}\label{first_step_rTT_ham}
    H_\gamma = \cosh ( \gamma ) H_0 - \sinh ( \gamma ) \sqrt{ H_0^2 - J_0^2 } \, ,
\end{align}
where $J_0$ is the conserved angular momentum in the seed theory. Deforming $H_\gamma$ by QM-$\TT$ using the general formula (\ref{QM_TT_deformed_hamiltonian}), we arrive at the first doubly-deformed Hamiltonian
\begin{align}\label{doubly_deformed_ham}
    H_{\gamma, \lambda} = \frac{1}{4 \lambda} \left( 1 - \sqrt{ 1 - 8 \lambda \left( \cosh ( \gamma ) H_0 - \sinh ( \gamma ) \sqrt{ H_0^2 - J_0^2 } \right) } \right) \, .
\end{align}
On the other hand, suppose that we first perform a QM-$\TT$ deformation of the harmonic oscillator to arrive at (\ref{QM_TT_deformed_hamiltonian}), where the initial condition is $H ( 0 ) = H_0$. In order to perform a subsequent QM-root-$\TT$ deformation, we first perform a Legendre transform of (\ref{QM_TT_deformed_hamiltonian}) to obtain the QM-$\TT$-deformed Lagrangian, which is
\begin{align}\label{TT-deformed-sho-lagrangian}
    L_\lambda &= \frac{1}{4 \lambda} \left( \sqrt{ \left( 1 + 4 \lambda \dot{x}^i \dot{x}_i \right) \left( 1 - 4 \lambda x^i x_i \right) } - 1  \right)  \, \nonumber \\
    &= \frac{1}{4 \lambda} \left( \sqrt{ \left( 1 + 4 \lambda X_2 \right) \left( 1 - 4 \lambda X_1 \right) } - 1  \right) \, ,
\end{align}
where in the last step we have written the Lagrangian in terms of the $SO(N)$ invariants defined in (\ref{QM_Xi_def}). The QM-root-$\TT$ operator constructed from this $L_\lambda$ is given by
\begin{align}\hspace{-20pt}
    R_\lambda = \sqrt{ \frac{1 + 2 \lambda \left( X_2 + 4 \lambda X_3^2 + 16 \lambda^2 X_1^2 X_2 - X_1 \left( 1 + 4 \lambda \left( X_2 + 4 \lambda X_3^2 \right) \right) \right) - \sqrt{ \left( 1 - 4 \lambda X_1 \right) \left( 1 + 4 \lambda X_2 \right) }}{8 \lambda^2 ( 1 + 4 \lambda X_2 )} } \, .
\end{align}
Therefore, to leading order in the QM-root-$\TT$ parameter $\gamma$, the Lagrangian obtained by first deforming the harmonic oscillator by QM-$\TT$ and then by QM-root-$\TT$ is 
\begin{align}\label{commute_L_lambda_gamma}
    L_{\lambda, \gamma} = L_\lambda + \gamma R_\lambda + \mathcal{O} ( \gamma^2 ) \, .
\end{align}
We claim that this disagrees with the Lagrangian $L_{\gamma, \lambda}$ which is obtained by Legendre transforming the Hamiltonian (\ref{doubly_deformed_ham}). Although it is unwieldy to peform the Legendre transform explicitly in closed form, it suffices to expand perturbatively in small momenta $p^i$ (or velocities $x^i$) and perform the Legendre transform to leading order in velocities. We will also specialize to the case $N = 2$ for simplicity.

To leading order in velocities and to first order in $\gamma$, the conjugate momenta associated with (\ref{doubly_deformed_ham}) are
\begin{align}
    p_1 &= \frac{1}{{ \left( x_1^2 + x_2^2 \right) \sqrt{1 - 4 \lambda \left( x_1^2 + x_2^2 \right) }  }} \Bigg[ \left( x_1^2 + x_2^2 \right) \dot{x}_1 + \gamma \left( x_1^2 - x_2^2 \right) \dot{x}_1 + 2 \gamma x_1 x_2 \dot{x}_2  \nonumber \\
    &\qquad - 2 \lambda \left( x_1^2 + x_2^2 \right) \left( x_2^2 \dot{x}_1 ( 2 - 3 \gamma ) + 4 x_1 x_2 \dot{x}_2 \gamma + x_1^2 \dot{x}_1 ( 2 + \gamma ) \right) \Bigg] + \mathcal{O} \left( \gamma^2, \dot{x}^2 \right) \, , \nonumber \\
    p_2 &= \frac{1}{{ \left( x_1^2 + x_2^2 \right) \sqrt{1 - 4 \lambda \left( x_1^2 + x_2^2 \right) }  }} \Bigg[  \left( x_1^2 + x_2^2 \right) \dot{x}_2 + \gamma \left( x_2^2 - x_1^2 \right) \dot{x}_2  + 2 \gamma x_1 x_2 \dot{x}_1 \nonumber \\
    &\qquad - 2 \lambda \left( x_1^2 + x_2^2 \right) \left( 4 \gamma x_1 x_2 \dot{x}_1 + ( 2 + \gamma ) x_2^2 \dot{x}_2 + ( 2 - 3 \gamma ) x_1^2 \dot{x}_2 \right) \Bigg] + \mathcal{O} \left( \gamma^2, \dot{x}^2 \right) \, .
\end{align}
Thus the perturbative Legendre transform of (\ref{doubly_deformed_ham}), keeping terms up to first order in $\gamma$ and up to second order in velocities, is given by
\begin{align}\label{L_gamma_lambda_answer}
    L_{\gamma, \lambda} &= p_1 \dot{x}_1 + p_2 \dot{x}_2 - H_{\gamma, \lambda} \nonumber \\
    &= \frac{1}{4 \lambda \left( x_1^2 + x_2^2 \right) \sqrt{ 1 - 4 \lambda ( x_1^2 + x_2^2 ) } } \Bigg\{  2 \lambda   x_1^4 \left(\gamma -2 \lambda  \left((\gamma +2)  \dot{x}_1^2+(2-3 \gamma )  \dot{x}_2^2\right)-2\right) \nonumber \\
    &\quad -32 \gamma  \lambda ^2  x_1^3  x_2  \dot{x}_1  \dot{x}_2+ x_1^2 \Big[ -\sqrt{1-4 \lambda  \left( x_1^2+ x_2^2\right)}+4 (\gamma -2) \lambda   x_2^2 \left(2 \lambda  \left( \dot{x}_1^2+ \dot{x}_2^2\right)+1\right) \nonumber \\
    &\quad +2 (\gamma +1) \lambda   \dot{x}_1^2-2 (\gamma -1) \lambda   \dot{x}_2^2+1 \Big] +8 \gamma  \lambda   x_1  x_2  \dot{x}_1  \dot{x}_2 \left(1-4 \lambda   x_2^2\right)  \nonumber \\
    &\quad + x_2^2 \Big[ -\sqrt{1-4 \lambda  \left( x_1^2+ x_2^2\right)}  + 2 \lambda  \Big( x_2^2 \left(\gamma  \left(6 \lambda   \dot{x}_1^2-2 \lambda   \dot{x}_2^2+1\right)-4 \lambda  \left( \dot{x}_1^2+ \dot{x}_2^2\right)-2\right)  \nonumber \\
    &\quad -(\gamma -1)  \dot{x}_1^2+(\gamma +1)  \dot{x}_2^2 \Big)+1 \Big] \Bigg\} + \mathcal{O} \left( \gamma^2, \dot{x}^4 \right) \, .
\end{align}
This expression is to be compared with $L_{\lambda, \gamma}$ in equation (\ref{commute_L_lambda_gamma}) which is obtained by performing the deformations in the other order. Expanding $L_{\lambda, \gamma}$ to second order in velocities in the special case $N=2$, and writing the results in terms of $x_1, x_2$, we find
\begin{align}\label{L_lambda_gamma_answer}
    L_{\lambda, \gamma} &= \frac{1}{4} \Bigg\{ \frac{\gamma  \sqrt{-4 \lambda  \left( x_1^2+ x_2^2\right)-2 \sqrt{1-4 \lambda  \left( x_1^2+ x_2^2\right)}+2}}{\lambda } + \frac{\sqrt{1-4 \lambda  \left( x_1^2+ x_2^2\right)}-1}{\lambda } \nonumber \\
    &\quad + \frac{\sqrt{2} \gamma}{\sqrt{1 -2 \lambda  \left( x_1^2+ x_2^2\right)-\sqrt{1-4 \lambda  \left( x_1^2+ x_2^2\right)}}} \Bigg[  \dot{x}_1^2 \sqrt{1-4 \lambda  \left( x_1^2+ x_2^2\right)} \nonumber \\
    &\quad + \left(4 \lambda  \left( x_1^2+ x_2^2\right)-1\right) \left(4 \lambda  ( x_2  \dot{x}_1- x_1  \dot{x}_2)^2+ \dot{x}_1^2+ \dot{x}_2^2\right)+ \dot{x}_2^2 \sqrt{1-4 \lambda  \left( x_1^2+ x_2^2\right)} \Bigg] \nonumber \\
    &\quad +2 \left( \dot{x}_1^2+ \dot{x}_2^2\right) \sqrt{1-4 \lambda  \left( x_1^2+ x_2^2\right)} \Bigg\} + \mathcal{O} \left( \gamma^2, \dot{x}^4 \right) \, .
\end{align}
Although the expressions for $L_{\gamma, \lambda}$ in (\ref{L_gamma_lambda_answer}) and $L_{\lambda, \gamma}$ in (\ref{L_lambda_gamma_answer}) are accurate only to leading order in $\gamma$ and to quadratic order in velocities, they retain terms of all orders in $\lambda$. It is already clear that the two Lagrangians are not equivalent because they have different $\lambda$ dependence. One can see this more explicitly by expanding the difference between the resulting Lagrangians at small $\lambda$:
\begin{align}\hspace{-20pt}
    L_{\gamma, \lambda} - L_{\lambda, \gamma} = \frac{\gamma \lambda}{2} \left( x_1^4 + 12 x_1 x_2 \dot{x}_1 \dot{x}_2 + 2 x_1^2 \left( x_2^2 + \dot{x}_1^2 - 2 \dot{x}_2^2 \right) + x_2^2 \left( x_2^2 - 4 \dot{x}_1^2 + 2 \dot{x}_2^2 \right) \right) + \mathcal{O} \left( \lambda^2, \gamma^2, \dot{x}^4 \right) \, .
\end{align}
Thus the difference between the two deformed Lagrangians vanishes when $\lambda = 0$, as it must, since in this case both Lagrangians are the result of performing only a QM-root-$\TT$ deformation by parameter $\gamma$. The difference also vanishes for $\gamma = 0$, since in this limit we have two theories that have only been deformed by QM-$\TT$. But when both $\lambda$ and $\gamma$ are non-zero, we see that the two Lagrangians differ at first order in $\lambda$ and $\gamma$.

In fact, even when all velocities are set to zero, we find 
\begin{align}
    L_{\gamma, \lambda} - L_{\lambda, \gamma} = \frac{\gamma \lambda}{2} \left( x_1^2 + x_2^2 \right)^2 + \mathcal{O} \left( \lambda^2, \gamma^2, \dot{x}^2 \right)
\end{align}
Therefore the two deformations do not even commute in the zero-momentum sector; the pure potential terms in the Lagrangians are sensitive to the order in which one performs the deformations.

As we have emphasized in Sections \ref{sec:bosonic} and \ref{sec:dim_red}, this non-commutativity is a feature of the $(0+1)$-dimensional flows but not of the parent $\TT$ and root-$\TT$ flows in two spacetime dimensions. Unlike the $2d$ case, in quantum mechanics it seems that there is some non-vanishing curvature in theory space along the directions whose tangent vectors are the QM-$\TT$ and QM-root-$\TT$ operators.

\bibliographystyle{utphys}
\bibliography{master}

\end{document}